\documentclass[sn-mathphys,Numbered]{sn-jnl}

\usepackage{graphicx}%
\usepackage{multirow}%
\usepackage{amsmath,amssymb,amsfonts}%
\usepackage{amsthm}%
\usepackage{mathrsfs}%
\usepackage[title]{appendix}%
\usepackage{xcolor}%
\usepackage{textcomp}%
\usepackage{manyfoot}%
\usepackage{booktabs}%
\usepackage{algorithm}%
\usepackage{algorithmicx}%
\usepackage{algpseudocode}%
\usepackage{listings}%
\usepackage{svg}
\usepackage{placeins}

\begin{document}

\title[Implementation of digital MemComputing using standard electronic components]{Implementation of digital MemComputing using standard electronic components}

\author*[1]{\fnm{Yuan-Hang} \sur{Zhang}}\email{yuz092@ucsd.edu}
\author*[1]{\fnm{Massimiliano} \sur{Di Ventra}}\email{diventra@physics.ucsd.edu}

\affil*[1]{\orgdiv{Department of Physics}, \orgname{Univerisity of California San Diego}, \orgaddress{\city{La Jolla}, \postcode{92093}, \state{CA}, \country{USA}}}

\abstract{Digital MemComputing machines (DMMs), which employ nonlinear dynamical systems with memory (time non-locality), have proven to be a robust and scalable  unconventional computing approach for solving a wide variety of combinatorial optimization problems. However, most of the research so far has focused on the numerical simulations of the equations of motion of DMMs. This inevitably subjects time to discretization, which brings its own (numerical) issues that would be otherwise absent in actual physical systems operating in continuous time. Although hardware realizations of DMMs have been previously suggested, their implementation would require materials and devices that are not so easy to integrate with traditional electronics. 
Addressing this, our study introduces a novel hardware design for DMMs, utilizing readily available electronic components. This approach not only significantly boosts computational speed compared to current models but also exhibits remarkable robustness against additive noise. Crucially, it circumvents the limitations imposed by numerical noise, ensuring enhanced stability and reliability during extended operations. This paves a new path for tackling increasingly complex problems, leveraging the inherent advantages of DMMs in a more practical and accessible framework.}


\maketitle

\section{Introduction}

Over the past half century, the rapid development of computers has mirrored the prediction of Moore's law, with the number of transistors doubling approximately every two years \cite{schaller1997moore}. As we near the physical limits of this growth, formidable challenges such as quantum tunneling, heat dissipation, and escalating production costs \cite{shalf2020future} necessitate a pivot towards novel computational paradigms. Unconventional computing, encompassing areas like quantum \cite{nielsen2010quantum}, neuromorphic \cite{Wong}, optical \cite{solli2015analog}, and molecular computing \cite{de2007molecular}, promises to not only sustain computational growth but also foster more energy-efficient and adaptable systems, capable of tackling problems currently beyond the reach of classical computers. 

One key idea in unconventional computing is to harness physical principles to greatly accelerate the solution of certain problems. For instance, an ideal (decoherence-free) quantum computer could in principle solve integer factorization in polynomial time utilizing quantum entanglement and interference \cite{shor1999polynomial}. A new class of machines, dubbed ``MemComputing''~\cite{diventra13a}, leverages time non-locality (memory) to solve a wide variety of computational problems efficiently \cite{traversa2015universal,di2022memcomputing}. In particular, their digital version (digital MemComputing machines or DMMs) have been designed to solve combinatorial optimization problems \cite{traversa2017polynomial}. 

We stress here that the prefix ``Mem'' stands for ``memory'', which is generally intended as ``time non-locality'', not necessarily storage~\cite{di2023memory}. Time non-locality is the {\it non-equilibrium} property of a physical system that when perturbed, the perturbation affects the system's state at a later time~\cite{di2022memcomputing}. Under appropriate conditions, it induces spatial non-locality~\cite{di2023memory}. This, in turn, initiates dynamical long-range order in the system \cite{di2017topological,di2019digital}, which is exploited by 
DMMs to scrutinize the structure of the target problem, thereby solving it efficiently. 

In the present paper we focus on DMMs \cite{traversa2017polynomial}, which have a finite set of input and output states that can be written/read with finite precision, hence are scalable to large problem sizes. The dynamics of a DMM are governed by a set of ordinary differential equations (ODEs) \cite{traversa2017polynomial,bearden2020efficient}, whose equilibria correspond to the solutions (if they exist) of the target problem. In continuous (physical) time, it has been proved that if such systems have equilibria (solutions), periodic orbits and chaos can be avoided~\cite{di2017chaos,di2017periodic}, and the convergence to the equilibrium point(s) can be reached with a number of jumps (more precisely, instantons, which are abrupt transitions in the trajectories, connecting critical points of the dynamics of different stability) that scales polynomially with problem size \cite{di2017topological,di2019digital}. 

Studies have illustrated that numerical simulations of DMM's equations can effectively solve a plethora of complex combinatorial optimization problems \cite{traversa2017polynomial,traversa2018evidence,traversa2018memcomputing,sheldon2019stress,bearden2020efficient}, significantly surpassing traditional algorithms in performance (see also industrial case studies performed at MemComputing, Inc.~\cite{Company}). Recently, it was numerically shown that MemComputing can solve the hardest integer factorization problems with quadratic complexity up to 300 bits \cite{sharp2023scaling}, with projections to factor 2048-bit RSA keys \cite{RSA} within weeks in software.

However, the numerical simulation of ODEs inevitably necessitates time discretization, leading to the accumulation and possible amplification of numerical errors over time \cite{stuart1994numerical}. Given that DMMs typically require extended simulation times to reach the solution of very large problems (in the hundreds of thousand or millions of variables), the numerical errors may ultimately cause the simulations of DMMs to fail \cite{zhang2021directed}. Although better numerical methods and careful control of such simulations can mitigate this issue, it comes with additional numerical overhead. 

The research gap, therefore, lies in overcoming these limitations inherent in numerical simulations of DMMs. A promising solution is the construction of a {\it physical} DMM that operates in continuous time.  Towards this objective, attempts have been made to realize DMMs using hardware components. For instance, Ref.~\cite{traversa2017polynomial} proposed a design employing resistive memories, but since these are not standard elements, their fabrication and integration into circuits could present a challenge. A similar issue pertains to the self-organizing logic gates of DMMs discussed in~\cite{Pinna}, where nanomagnets have been suggested as possible memory materials. Alternatively, Ref.~\cite{nguyen2023hardware} realized a DMM in hardware using field-programmable gate arrays, which achieved considerable speed-up compared to simulations. However, this implementation is still based on the discretization of time. As such, it essentially falls under the category of numerical simulations (albeit directly in hardware), implying that the previously mentioned challenges persist.

Addressing this gap, our work aims to design a physical, hardware-based DMM that operates in continuous time, exclusively employing {\it standard} electronic components. The strategy is to start from a set of ODEs describing a DMM solving, e.g., a combinatorial problem, and then look for  standard hardware components that would reproduce such dynamics. As a prototypical example we consider the Boolean satisfiability (SAT) problem with three literals per clause. Any other combinatorial problem can, in principle, be mapped into it \cite{cook1971complexity}.

Given that we are essentially conducting hardware-based ODE simulations, our design concept draws inspiration from analog computers prevalent in the 1960s, predominantly utilizing operational amplifier (op-amp)-based circuits to perform mathematical operations. 

However, a key issue still persists: by moving from software to hardware we have exchanged numerical noise for physical noise. Analog computers largely fell out of favor in the 1970s in part due precisely to physical noise, which limited their scalability. Factors like component variability in resistors and capacitors, temperature sensitivity, parasitic effects, and jitter all contribute to this physical noise, compromising the accuracy of analog signals. While careful design can mitigate some of these noise sources, they can never be completely eliminated.

While physical noise is detrimental for analog computers, they do not pose much of a problem for DMMs. This is because, just like modern digital computers, despite the use of physical signals, a DMM is a {\it digital} machine, since the input and output states of a DMM are finite and can be read/written with {\it finite precision}. Moreover, it was demonstrated that, the transition function between the input and the output states of a DMM is of topological character, making them robust to perturbative noise \cite{di2017topological,di2019digital}. This is key for their scalability as a function of problem size, a feature that is not shared by analog machines. In fact, it has also been shown that introducing some (Gaussian) physical noise can, in certain cases, even aid the solution \cite{primosch2023self}.

One additional key distinction between physical and numerical noise is worth noting: physical noise is typically localized in space and time and does not accumulate over extended periods~\cite{di2022memcomputing}. In contrast, numerical noise can accumulate and may even amplify exponentially over time~\cite{stuart1994numerical}. Therefore, for large-scale, long-time simulations, a physical DMM, even with reasonably high noise levels, is expected to eventually outperform numerical simulations.

In this paper, we outline a prototype of such a physical DMM. We are aware that some of the design elements we have used may not be optimal. However, our goal was not to achieve optimality in design. Rather, we hope that our work would serve as a foundational demonstration that can inspire future developments.

This paper is organized as follows. In Sec.~\ref{sec:mem}, we outline the formulation of the DMM equations and introduce a circuit that implements these equations. Results from hardware emulations using LTspice and numerical simulations with Python are presented in Section~\ref{sec:result}, and noise resistance of our model is also demonstrated. We conclude the paper with a few remarks in Sec.~\ref{sec:conclusion}. The technical details necessary to reproduce our paper can be found in the Appendix. 

\section{Solving 3-SAT with MemComputing}
\label{sec:mem}
 In a SAT problem, the task is to determine whether there exists an assignment of truth values to a set of Boolean variables, such that the given Boolean equation evaluates to true. Despite its simple formulation, the solution of SAT problems is often required in various industrial applications such as circuit design, logistics, scheduling, etc. \cite{eggersgluss2010robust,gomes2000heavy}, and an efficient solver is desirable. In particular, we focus on the three-satisfiability (3-SAT) problem, where the Boolean equation is a conjuction of clauses, each of which is a disjuction of three literals, where a literal is either a Boolean variable or its negation.

\subsection{MemComputing equations for 3-SAT}
The approach for solving 3-SAT problems using DMMs was already reported in Ref.~\cite{bearden2020efficient}, where the dynamics of the DMM was written as a set of ordinary differential equations (ODEs) and solved numerically. For an easier hardware implementation, we slightly modified the equations used in Ref.~\cite{bearden2020efficient}. For a 3-SAT problem with $N$ variables and $M$ clauses, the resulting equations are presented below: 

\begin{align}
\dot{v}_n=&\sum_m\bigg( \eta\,\text{softmax}(\mathbf{x}_{l}^n)_mx_{s,m}G_{n,m}(v_n,v_j,v_k)\label{eq:v}\\\nonumber
+&(1+\zeta\eta\,\text{softmax}(\mathbf{x}_{l}^n)_m)\left(1-x_{s,m}\right) R_{n,m}(v_n,v_m,v_k)\bigg)\\
\dot{x}_{s,m}=&\beta \left( x_{s,m} +\epsilon \right)\left( C_m(v_i,v_j,v_k)-\gamma\right),\label{eq:xs}\\
\dot{x}_{l,m}=&\alpha e^{-x_{l, m}} \left( C_m(v_i,v_j,v_k)-\delta\right),\label{eq:xl}\\
C_m =& 1 - \max(\tilde{v}_{i,m}, \tilde{v}_{j,m}, \tilde{v}_{k,m}),\label{eq:C}\\
G_{n,m}=&q_{n,m}C_m(v_n, v_j, v_k),\label{eq:G}\\
R_{n,m}=&\begin{cases}
   q_{n,m}C_m(v_n, v_j, v_k), \\
    \hspace{1cm} \text{if } C_m(v_n,v_j,v_k)=1-q_{n,m}v_n ,\\
    0, \hspace{7mm} \text{otherwise}.
  \end{cases}\label{eq:R}\\
\mathrm{softmax}(\mathbf{z})_i=&\frac{e^{z_i}}{\sum_j e^{z_j}} \label{eq:softmax}
\end{align}

Here, $v_n\in[0, 1]$ represent the continuously relaxed Boolean variables in the 3-SAT problem ($n=1, \cdots, N$), while $x_{l,m}\in[0, M]$ and $x_{s, m}\in[0, 1]$ are the long-term and short-term memory variables, respectively ($m=1, \cdots, M$). Whenever a variable exceeds the bounds, it is reset to the bound value, which is handled by the circuit detailed in the next section. $\alpha, \beta, \gamma, \delta, \epsilon, \zeta$ and $\eta$ are constants. 

Each clause is represented by the clause function, $C_m(v_i, v_j, v_k)$, indicating that the $i$-th, $j$-th and $k$-th Boolean variables are present in the $m$-th clause. If the $i$-th literal in the $m$-th clause is negated, then $q_{i, m}=-1$ and $\tilde{v}_{i,m}=1-v_i$, otherwise $q_{i,m}=1$ and $\tilde{v}_{i,m}=v_i$. The definition of $C_m$ in Eq.~\eqref{eq:C} gives it the range [0, 1], and the clause is closer to satisfaction if $C_m$ is closer to 0. 

The gradient-like term, $G_{n, m}$ (Eq.~\eqref{eq:G}), aims to push all literals in a clause towards satisfaction uniformly, while the rigidity term, $R(n, m)$ (Eq.~\eqref{eq:R}), tries to hold the literal closest to being satisfied in place. The short-term memory, $x_{s, m}$, functions to alternate between the gradient-like and the rigidity terms, while the long-term memory, $x_{l, m}$, assigns a dynamic weight to each of the clauses. A more comprehensive interpretation of each of these parameters and terms can be found in \cite{di2022memcomputing,bearden2020efficient}. Here, the parameters are chosen to be $\alpha=5, \beta=20, \gamma=0.25, \delta=0.05, \epsilon=10^{-3}$, and $\eta=3000$. The optimal $\zeta$ decreases according to a power law in a logarithmic scale with system size $N$, and the details can be found in Appendix \ref{appendix:param}.

Comparing to the original equations in Ref.~\cite{bearden2020efficient}, we adjusted Eq.~\eqref{eq:v} by rescaling and computing the softmax of the long-term memories $\mathbf{x}_l$. This bounds the range of the multiplier, making the circuit implementation easier. 

\subsection{Hardware implementation of the equations}
Eqs.~\eqref{eq:v}-\eqref{eq:R} contain basic arithmetic operations, exponentiation, maximum and softmax functions. Linear operations can be implemented using op-amp based arithmetic circuits, while the exponential function can be realized using the exponential relation between a bipolar junction transistor's emitter current and base voltage. We will now illustrate how to realize these ODEs using these basic components, with the detailed design of each operation provided in \ref{appendix:circuit}.

The variables $v_n$, $x_{s,m}$ and $x_{l,m}$ are represented as voltages across capacitors, and their time derivatives are simulated using currents charging and discharging the capacitors. 

\begin{figure}[htbp]
	\centering
	\includegraphics[width = 0.6\textwidth]{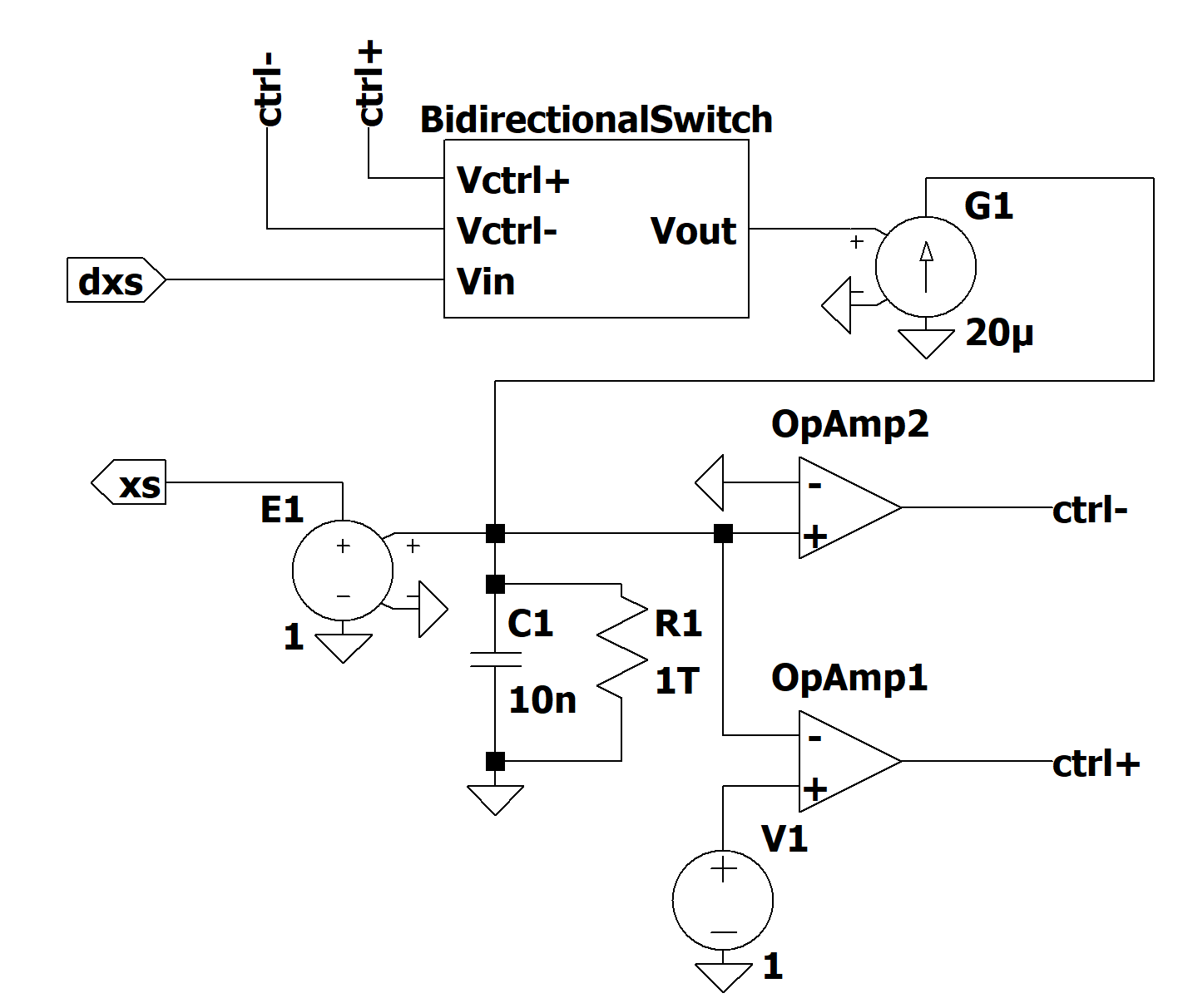}
	\caption{Circuit for the implementation of the short-term memory variable, $x_s$. The circuit takes in the pre-computed voltage, $dx_s$, integrates it, and subsequently produces the updated value of $x_s$.}
	\label{fig:xs}
\end{figure}

Fig.~\ref{fig:xs} presents the circuit implementing the short-term memory variable, $x_s$. The value of $x_s$ is represented by the voltage across a 10 nF capacitor, $C_1$. A very large resistor, $R_1$, is connected in parallel to $C_1$ for stability, and its impact on the dynamics is negligible. A voltage-controlled current source, $G_1$, charges $C_1$. The control voltage, $dx_s$, which is the right-hand side of Eq.~\eqref{eq:xs}, is pre-computed in another segment of the circuit.

Before controlling the current source $G_1$, $dx_s$ is first passed into a bidirectional switch regulated by two control signals, $\text{ctrl}_+$ and $\text{ctrl}_-$. Each of these signals selectively allows or blocks input signals of a specific polarity. As $x_s$ is confined within the [0, 1] range, we use two open-loop operational amplifiers functioning as comparators to maintain these bounds, generating the control signals. Specifically, $\text{ctrl}_\pm$ equals 5V when the voltage across $C_1$ falls within the [0V, 1V] range and -5V when the voltage surpasses its respective limit.

In operation, when the voltage representing $x_s$ remains within the bounds (0V to 1V), the bidirectional switch allows the input signal, $dx_s$, to pass through. This signal influences the current source $G_1$, which then charges or discharges the capacitor $C_1$ as per Eq.~\eqref{eq:xs}. If $x_s$ exceeds 1V, the control signal $\text{ctrl}_+$ becomes negative, prompting the bidirectional switch to block positive $dx_s$ signals, thus allowing only negative $dx_s$ to pass. This mechanism enforces the upper boundary for $x_s$. The same process, in reverse, handles any voltage drops below 0V to uphold the lower boundary.

To maintain the capacitor's charge and avoid any additional input or output current to $C_1$, we use a voltage-controlled voltage source with unity gain, $E_1$, for isolation and to output the value of $x_s$. This isolation can be practically achieved through a voltage follower circuit made with a single op-amp.

\begin{figure}[htbp]
	\centering
	\includegraphics[width = 0.7\textwidth]{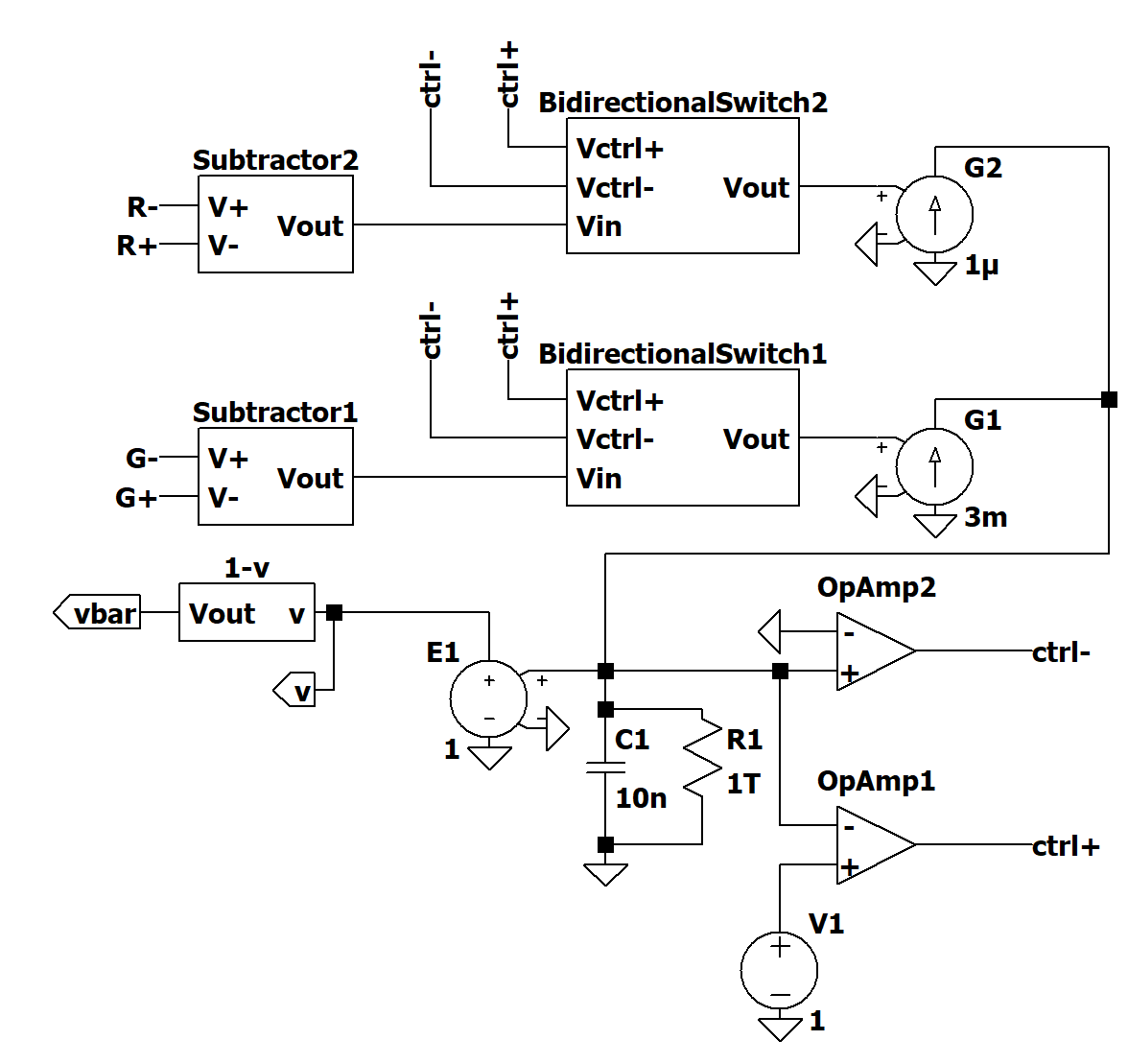}
	\caption{Implementation of the voltage dynamics, Eq.~\eqref{eq:v}, with a built-in negation mechanism. }
	\label{fig:v}
\end{figure}

To implement Eq.~\eqref{eq:v} for the variables $v_n$, we use a similar setup, depicted in Fig.~\ref{fig:v}. As each literal in 3-SAT can be either negated or not, this circuit simultaneously generates two values, $v$ and $\bar{v}=1-v$. Likewise, inputs are also grouped by their polarities (the sign mismatches in Fig.~\ref{fig:v} result from the inverted adders in the previous step). Although the constant $\eta=3000$ in Eq.~\eqref{eq:v} (represented by $G_1=3\text{mA}/\text{V}$ in Fig.~\ref{fig:v}) renders the rigidity term without it seemingly negligible (represented by $G_2=1\mu\text{A}/\text{V}$ in Fig.~\ref{fig:v}), numerical experiments corroborate that the inclusion of a small rigidity term aids in convergence towards the solution.

Implementing the long-term memory dynamics accurately, as per Eq.~\eqref{eq:xl}, requires more consideration due to the variability of the exponential term across several orders of magnitude. To address this, we choose to work in the logarithmic space and rewrite Eq.~\eqref{eq:xl} as:
\begin{equation}
    \dot{x}_{l,m}=\alpha \left(\exp(\log(C_m+\lambda)-x_l) - \exp(\log(\delta+\lambda)-x_l)\right).\label{eq:xl_expanded}
\end{equation}
Given that $C_m-\delta$ can be either positive or negative, taking the logarithm directly would pose issues. Therefore, we split it into two terms, $C_m+\lambda$ and $\delta+\lambda$, with $\lambda=0.1$ to ensure both terms are always positive. We then evaluate each product separately using log-sum-exp computations, and compute their difference to restore Eq.~\eqref{eq:xl}. Necessary constants are incorporated to rescale the signals to suitable ranges, and the details can be found in \ref{appendix:circuit}. 

\begin{figure}[htbp]
	\centering
	\includegraphics[width = 0.7\textwidth]{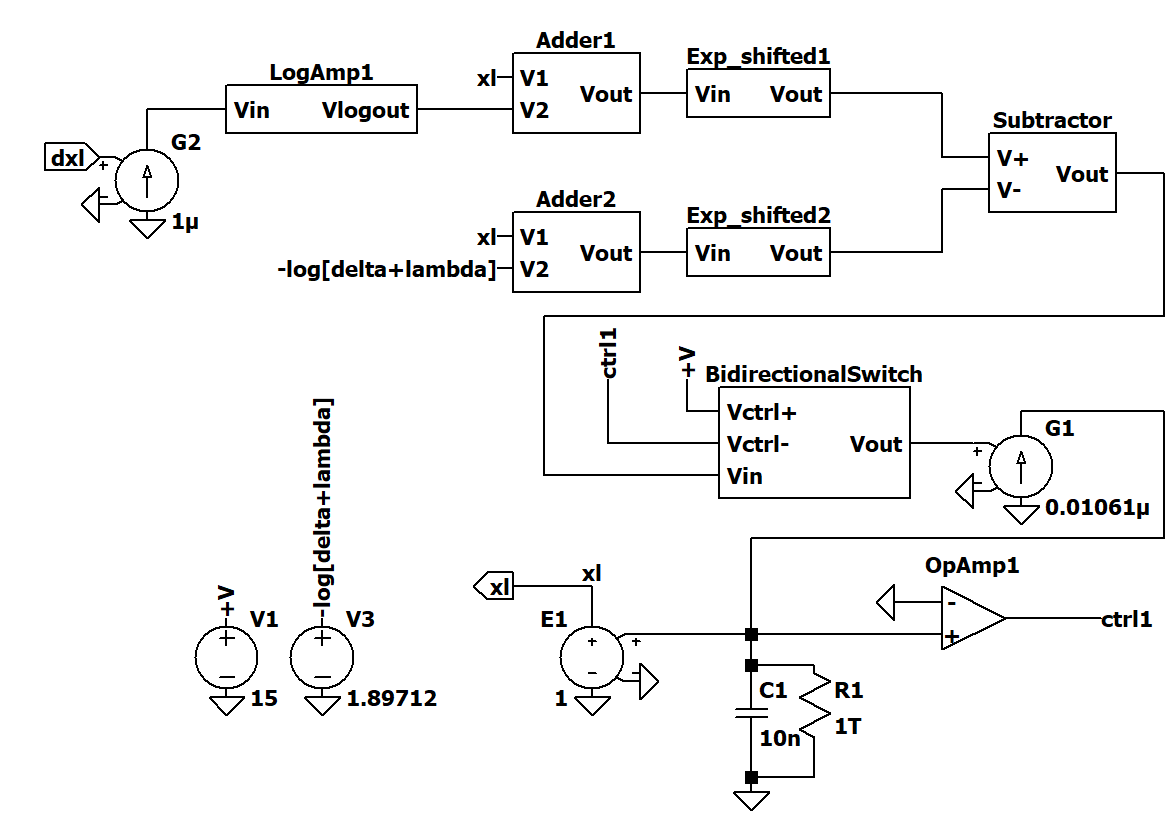}
	\caption{Implementation of the long-term memory dynamics, Eq.~\eqref{eq:xl}. Here, the input $dx_l=C_m+\lambda$, and log-sum-exp computation is employed to restore Eq.~\eqref{eq:xl}. Note that our implementation of the logarithm and exponential amplifiers contains a negative sign (see \ref{appendix:circuit}). }
	\label{fig:xl}
\end{figure}

\begin{figure}[htbp]
	\centering
	\includegraphics[width = 0.8\textwidth]{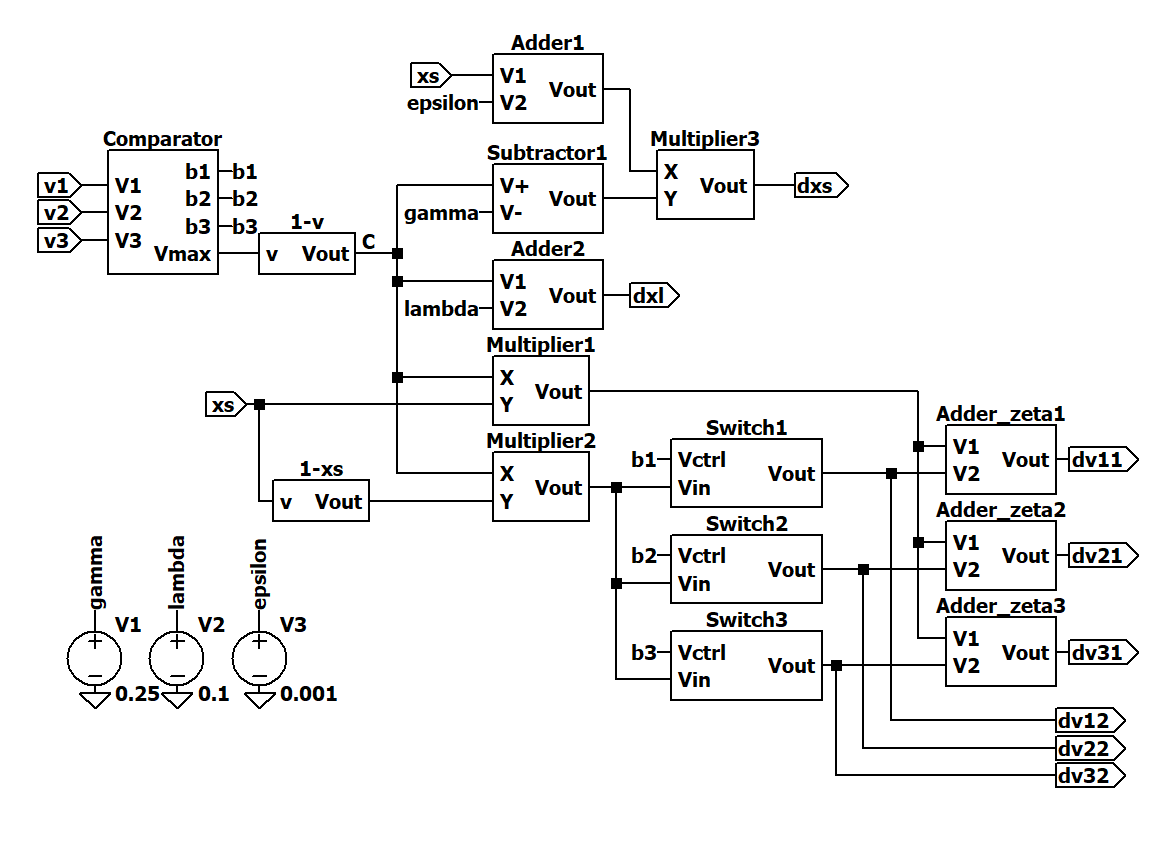}
	\caption{The module computing the time derivatives of the variables. It accepts the possibly negated variables, $v_1, v_2$ and $v_3$, and the short-term memory $x_s$, and outputs the intermediate results of their time derivatives, $dx_s$, $dx_l$, $dv_{n,1}$ and $dv_{n,2}$. }
	\label{fig:SOLG}
\end{figure}

Finally, Fig.~\ref{fig:SOLG} shows the evaluation of the clause function, gradient-like and rigidity terms, and how they contribute to the computation of the time derivatives in Eqs.~\eqref{eq:v}-\eqref{eq:xl}. The comparator module initially calculates the maximum of the three input (possibly negated) voltages, $v_{\max}$, along with three control signals, $b_1, b_2$ and $b_3$, indicating which input voltage attains the maximum value. The clause function is then computed as $C_m=1-v_{\max}$, which is further employed to evaluate $dx_s$ and $dx_l$ according to Eq.~\eqref{eq:xs} and Eq.~\eqref{eq:xl_expanded}. The rigidity term is calculated with the assistance of the control signals $b_i$, setting the specified values to 0. Finally, the right-hand-side of Eq.~\eqref{eq:v} is divided into two parts, $dv_{n,1}=x_{s,m}G_{n,m} + \zeta(1-x_{s,m})R_{n, m}$, which is later multiplied with the softmax module (Fig.~\ref{fig:softmax}); and $dv_{n,2}=(1-x_s)R_{n, m}$, which is added at the end. This final processing step is not depicted in Fig.~\ref{fig:SOLG}. 

By integrating all the modules in accordance with the topology of the target 3-SAT problem, we successfully develop a continuous-time, hardware-accelerated DMM. We stress here that in large-scale applications, the topology, or connectivity, of the 3-SAT problem (and its conjunctive normal form representation), can be programmed using cross-point switches or field-programmable gate arrays. This would allow switching from one type of problem to another with the same type of hardware design. 
Given our selection of capacitance and current source, it can be verified that one second of circuit dynamics equates approximately 100 units of time in Eqs.~\eqref{eq:v}-\eqref{eq:R}. With high-speed op-amps, it is possible to further reduce the characteristic time scale of the circuit to accelerate the solution. Limited by the numerical accuracy and convergence in our circuit emulations, we leave this as a future work.

\section{Results}
\label{sec:result}
To investigate the accuracy and efficiency of our methods, we emulated Eqs.~\eqref{eq:v}-\eqref{eq:R} in LTspice \cite{vladimirescu1994spice,kundert2006designer}  using the realistic circuit elements as we have discussed in the preceeding Sec.~\ref{sec:mem}. Furthermore, we also developed a Python program to simulate the same ODEs directly using numerical solvers. The full model files are available at \cite{github_link}.

To generate challenging 3-SAT instances, we utilized the unbiased generator with a planted solution as proposed in \cite{barthel2002hiding}, specifically targeting instances near the complexity peak at a clause-to-variable ratio of 4.3.

\begin{figure}[htbp]
	\centering
	\includegraphics[width = 0.7\textwidth]{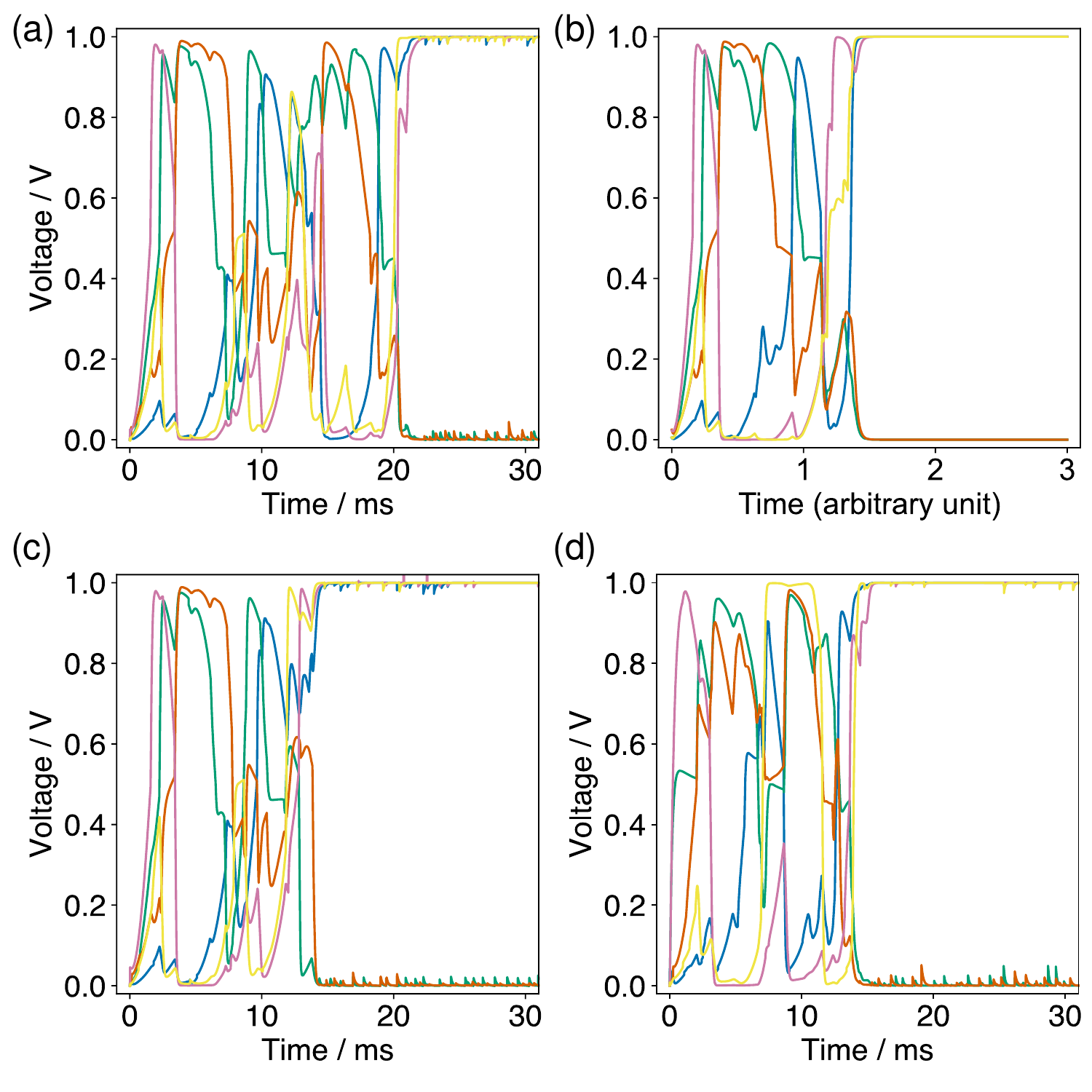}
	\caption{Comparison of the trajectories of the first five variables in a 3-SAT problem with 10 variables and 43 clauses for (a) LTspice circuit emulation, with no noise added, (b) Python numerical simulation, (c) LTspice circuit emulation, with $10\%$ white noise, and (d) LTspice circuit emulation, with $20\%$ white noise. With the same initial conditions, the trajectories are initially similar, but soon differ due to differences in noise (both physical and numerical). However, despite the paths are different, eventually they all converge to the same solution. }
	\label{fig:traj}
\end{figure}

As a proof-of-concept test, we compared both methodologies using a 3-SAT problem consisting of 10 variables and 43 clauses. Fig.~\ref{fig:traj}(a)(b) illustrates the trajectories of the first five variables. Both simulations started with identical initial conditions, and we can see that the initial part of their trajectories are the same. However, due to noise and minor discrepancies in different implementations, their trajectories soon differ. Nevertheless, despite following completely different trajectories, both simulations promptly converged to the same solution. 

To better understand the noise sensitivity of our design, we introduced white noise of varying strength to all relevant voltage sources in LTspice emulations, except for those where noise is unlikely to make a significant impact (e.g., the power supplies for op-amps). Of course, this added noise compounds the numerical one which is intrinsic in the discretization of time of the simulations. The findings are displayed in Fig.~\ref{fig:traj}(c)-(d). In these figures, the added noise strength is set at $10\%$ of the corresponding voltage for (c), and $20\%$ for (d). Again, we see that despite different trajectories, all simulations converge to the same solution, regardless of noise (both physical or numerical).

This example serves to illustrate the inherent robustness of DMMs against noise~\cite{di2022memcomputing}. To understand this robustness, note that, with the values of the voltage sources entering computations via arithmetic circuits, the injected white noise can affect the parameters $\gamma, \delta, \epsilon$ in Eqs.~\eqref{eq:v}-\eqref{eq:R}, but cannot change the equations' functional form. According to our analysis in Appendix \ref{appendix:param}, although an optimal set of parameters exist, the DMM still works consistently for a wide range of parameters and always converge to the solution if the parameters do not deviate too much from their optimal values.   

In fact, as observed in Fig.~\ref{fig:traj}, minor perturbations can cause the trajectory to diverge significantly. 
Yet, all paths ultimately converge towards the solution of the problem, demonstrating the model's resilience. Furthermore, this example also affirms the functional equivalence between the physical circuit built using realistic elements and the numerical Python-based implementation.

To extend the scope of our assessment, we generated additional 3-SAT instances with varying problem sizes and solved them through the DMM circuit simulation. Given the substantial resource consumption of realistic circuit emulations in LTspice, we opted for the Python simulation in this test, which allowed us to simulate much larger systems.

\begin{figure}[htbp]
	\centering
	\includegraphics[width = 0.7\textwidth]{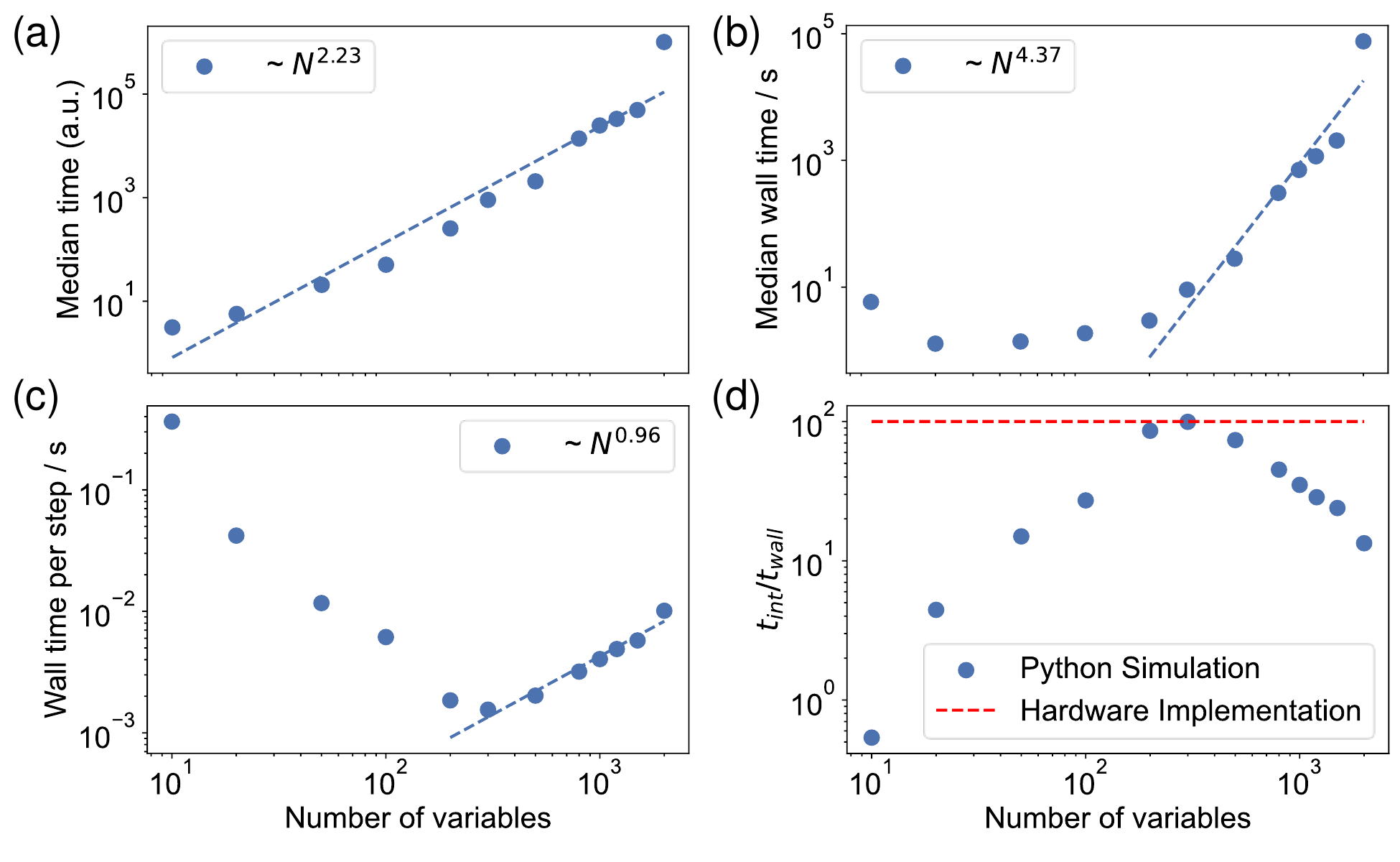}
	\caption{  (a) Median integration time (in arbitrary units) as a function of the number of variables in the 3-SAT problem, which were generated near the complexity peak at a clause-to-variable ratio of 4.3. The curve approximates a power-law fit, yielding an exponent of $2.23\pm 0.17$. (b) The progression of the median wall time for solving problems, exhibiting near-constant times for smaller problems and scaling approximately as $N^{4.37\pm 0.55}$ for larger problems. (c) Wall time per integration step. Excluding the initialization time, this grows linearly with the system size. (d) The ratio of integrated time to real-world seconds, peaking around 100 for the Python simulation, and declining as the system size expands. Excluding the smaller cases where the Python simulation's initialization time is dominant, our hardware-accelerated DMM would achieve up to an $8\times$ speed increase compared to the numerical simulations.}
	\label{fig:time}
\end{figure}

Fig.~\ref{fig:time}(a) illustrates how the median integration time (in arbitrary units) scales with the number of variables in the 3-SAT problem. These problems were generated near the complexity peak with a clause-to-variable ratio of 4.3. Each data point was determined by solving 100 distinct 3-SAT instances, with the median being recorded once 51 of them were solved. A power-law fit of the median solution time against the number of variables yielded an exponent of $2.23\pm 0.17$. This result marginally outperforms the findings reported in \cite{bearden2020efficient}.

Similarly, Fig.~\ref{fig:time}(b) displays the median wall time, where 100 instances are simulated in parallel on a single NVIDIA TITAN RTX GPU, and Fig.~\ref{fig:time}(c) shows the average wall time per integration step. For smaller cases, the program initialization time dominates, resulting in a nearly constant wall time and a greater wall time per step. However, as the number of variables grows, the initialization time becomes insignificant, and the wall time per step increases linearly with the problem size.

Finally, Fig.~\ref{fig:time}(d) computes the ratio between the median integration time and the median wall time, which indicates how many units of time we can integrate per real-world second. In our previously detailed hardware implementation, one second corresponds to approximately 100 units of time (as represented by the red dotted line). Compared to this, the ratio in the Python simulation reaches a peak of around 100 and gradually diminishes as the system size expands. If we assume the hardware solution shares the same scaling properties of the numerical simulation, this means that for a 3-SAT problem with 2000 variables, our circuit design would achieve an 8-fold acceleration compared to the numerical simulations.

However, we expect the hardware solution to scale better compared to numerical simulations~\cite{di2022memcomputing}. In the Python simulations presented above, the discretized time step $\Delta t$ is of the order of $10^{-1}$. For more challenging problems, $\Delta t$ needs to be reduced, resulting in an increase in the number of integration steps needed \cite{bearden2020efficient}. However, our envisioned hardware design, which operates in continuous time, integrates the ODEs at a consistent rate, regardless of the problem size or difficulty. This intrinsic characteristic of the hardware, once practically implemented and verified, is expected to lead to substantial speed improvements. 

Moreover, since numerical noise is eliminated in hardware, we expect this approach to offer enhanced stability over extended periods of dynamics. This is another key advantage of our hardware-based approach over numerical simulations.

Despite promising results depicted in Fig.~\ref{fig:time}, the initial Python simulations did not account for physical imperfections that are typically present in realistic settings, such as component tolerance, capacitor leakage, op-amp input offset voltage, and temperature variations. These factors are crucial as they can significantly influence the dynamics and reliability of our system.

Among these factors, the tolerance of resistors stands out as potentially the most impactful on the system dynamics. Consider, for example, the adder circuit shown in Fig.~\ref{fig:adder}, which is governed by the equation:

\begin{equation}
    V_\mathrm{out}=\frac{V_1 R_2 + V_2 R_1}{R_1+R_2}\cdot\frac{R_3+R_4}{R_4}.
\end{equation}

With $R_1=R_2=R_3=R_4$, we have

\begin{equation}
    \delta V_\mathrm{out}=\frac{1}{2}V_1\left(-\frac{\delta R_1}{R_1}+\frac{\delta R_2}{R_2}+\frac{\delta R_3}{R_3}+\frac{\delta R_4}{R_4}\right) + \frac{1}{2}V_2\left(\frac{\delta R_1}{R_1}-\frac{\delta R_2}{R_2}+\frac{\delta R_3}{R_3}+\frac{\delta R_4}{R_4}\right).\label{eq:tol}
\end{equation}

In practical terms, if each resistor has a tolerance of $1\%$, this would result in an approximate $2\%$ error in the computed value of $V_\mathrm{out}$. Such errors are likely to accumulate through cascaded addition and multiplication operations, significantly magnifying deviations from the ideal dynamics.

On the other hand, temperature variations affect the thermal voltage, $V_T=k_BT/q$, which influences the characteristics of transistors crucial to the design of logarithmic amplifiers and multipliers. Appropriate designs incorporate temperature compensations based on paired transistors, cancelling out the impact of temperature fluctuations across a broad range of operating temperatures, rendering it negligible compared to the influence of resistor tolerance. Nonetheless, the error induced by temperature variations is multiplicative, similar to that caused by resistor tolerance, and we model them together, as detailed below. 

Comparatively, factors such as capacitor leakage and op-amp input offset voltage present less risk to the system's dynamics. For instance, capacitor leakage currents are generally around $10^{-2}\mu\mathrm{A}/(\mu\mathrm{F}\cdot \mathrm{V})\cdot CV$ for aluminum electrolytic capacitors \cite{IEC60384-4}. Similarly, typical input offset voltages for op-amps are on the order of $100\mu$V. These values are relatively insignificant when compared to the operational voltages and currents in our system, thereby having a lesser impact on the system's performance.

To validate our system's resilience against physical imperfections, we incorporated simulations of component tolerance and capacitor leakage into our Python model. We modeled resistor tolerance and temperature variation together as a multiplicative noise, $\eta$, affecting each addition and multiplication operation described in Eqs.~\eqref{eq:v}-\eqref{eq:R}. Since resistor tolerance is dominant, based on Eq.~\eqref{eq:tol}, $\eta$ is approximately twice the actual resistor tolerance. Since an error $\eta$ is present in every addition and multiplication operation, $\eta=1\%$ will result in an estimated $5\%$ error in the computed derivatives.

Capacitor leakage was integrated as an additional term in Eqs.~\eqref{eq:v}-\eqref{eq:xl}, modifying the derivatives according to:
\begin{equation}
    \tilde{\dot{v}}_n=\dot{v}_n-\kappa v_n
\end{equation}
and similarly for $x_l$ and $x_s$. We selected $\kappa=10^{-3}$, representing a leakage current $I_\mathrm{leak}=10^{-1}\mu\mathrm{A}/(\mu\mathrm{F}\cdot \mathrm{V})\cdot CV$, which is ten times the standard rate for aluminum electrolytic capacitors \cite{IEC60384-4}. 

\begin{figure}[htbp]
	\centering
	\includegraphics[width = 0.6\textwidth]{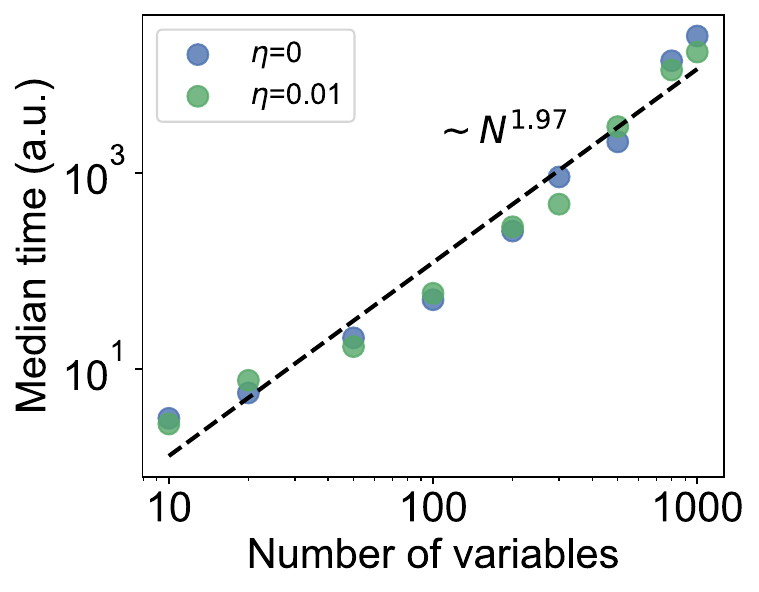}
	\caption{Median integration time (in arbitrary units) as a function of the number of variables, $N$, for 3-SAT problems at clause-to-variable ratio 4.3. This simulation incorporates an error $\eta$ in every addition and multiplication operation, attributable to resistor tolerances, and a capacitor leakage current $I_\mathrm{leak}=10^{-1}\mu\mathrm{A}/(\mu\mathrm{F}\cdot \mathrm{V})\cdot CV$. For each data point, at least 51 out of 100 distinct 3-SAT instances are solved to calculate the median integration time. The simulation here indicates that capacitor leakage and a small resistor tolerance has a minimal impact on the dynamics, and the median integration time scales essentially quadratically with the problem size. }
	\label{fig:tolerance}
\end{figure}

The simulation results for low noise levels, $\eta = 0$ and 0.01, are shown in Fig.~\ref{fig:tolerance}. We observe that such minimal noise levels do not impair the solution capabilities of our system, which continues to solve hard 3-SAT problems effectively. Importantly, the median time-to-solution scales essentially quadratically with the problem size, indicating robust performance under these conditions.

In Appendix~\ref{appendix:tolerance}, we present the complete scaling curves for $\eta$ values up to 0.2 and variable counts ($N$) up to 1500. The results demonstrate that our system maintains reliable performance in solving hard 3-SAT problems for $N \leq 500$, even at higher noise levels. However, as expected, beyond this threshold, the median solution time starts to exhibit exponential growth for much larger $\eta$. These findings highlight the robustness of our system’s design, even under moderate noise conditions.


\section{Conclusion}
\label{sec:conclusion}

In this paper, we have introduced a hardware design of a DMM that leverages only standard electronic components, which can be readily built using available technology, without the need of special materials or devices. By means of both LTspice and Python simulations, we have validated the reliability of our approach in tackling difficult 3-SAT problems even in the presence of physical noise under realistic conditions. Operating in continuous time, this hardware methodology completely removes numerical noise, aligning closely with the {\it physical} DMM concept, which has been suggested as a powerful alternative to solve hard combinatorial optimization problems \cite{traversa2017polynomial}. 

In finalizing our study, it is crucial to acknowledge the potential disparities between our simulated design and its real-world hardware implementation. Factors like parasitic effects at high frequencies and the distinct characteristics of physical noise all pose challenges often unaccounted for in simulations. These elements highlight the imperative for rigorous experimental validation and adjustments to ensure the practical robustness and performance of our circuit design. Moving forward, our future research will focus on the experimental implementation of this design, aiming to bridge the gap between simulations and real continuous-time dynamics. 


\section*{Acknowledgements} 
This work was supported by the National Science Foundation grant No. ECCS-2229880. 

M.D. is the co-founder of MemComputing, Inc. (\url{https://memcpu.com/}) that is attempting to commercialize the memcomputing technology. All other authors declare no competing interests.

All code and model files mentioned in the paper can be found at \cite{github_link}. 

\begin{appendices}

\section{Details of the circuit design}
\label{appendix:circuit}
In this Appendix, we present details of the hardware design to realize Eqs.~\eqref{eq:v}-\eqref{eq:R}, illustrating the individual building blocks of the circuit that performs different arithmetic functions. 

\begin{figure}[htbp]
	\centering
	\includegraphics[width = 0.45\textwidth]{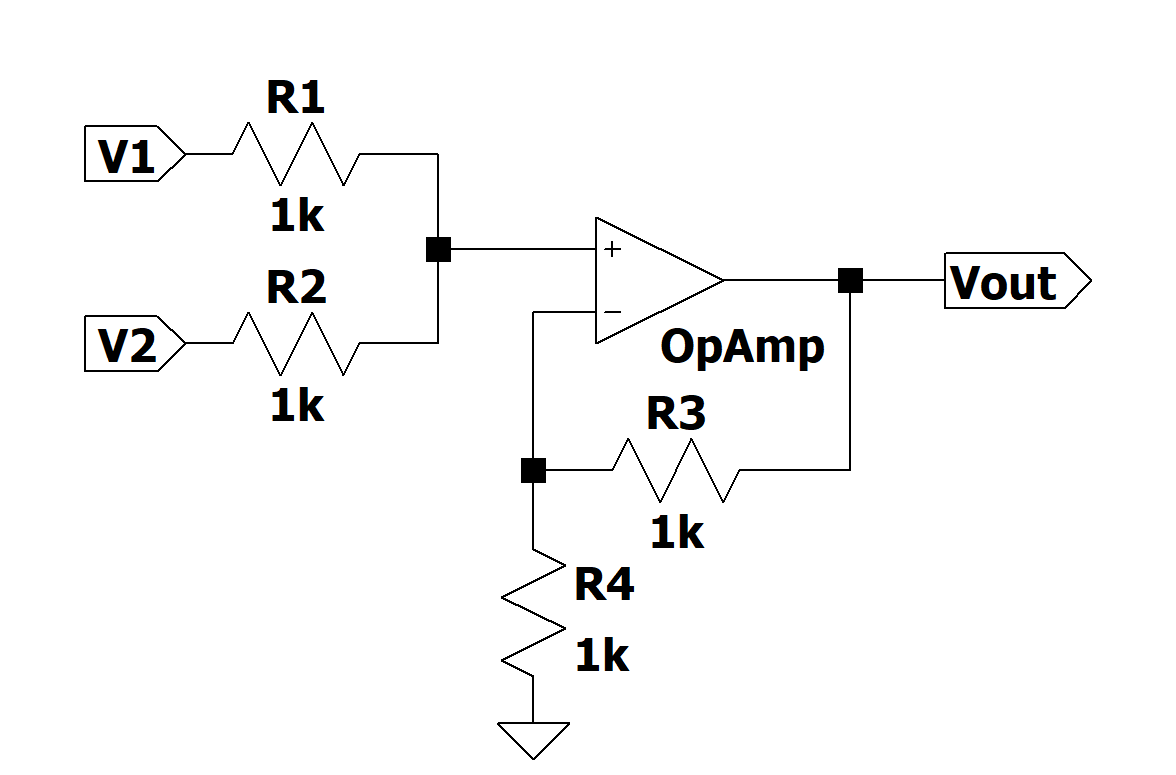}
	\caption{The adder circuit, realized using an op-amp feedback loop. It computes the function $V_{\text{out}}=V_1+V_2$.}
	\label{fig:adder}
\end{figure}

\begin{figure}[htbp]
	\centering
	\includegraphics[width = 0.45\textwidth]{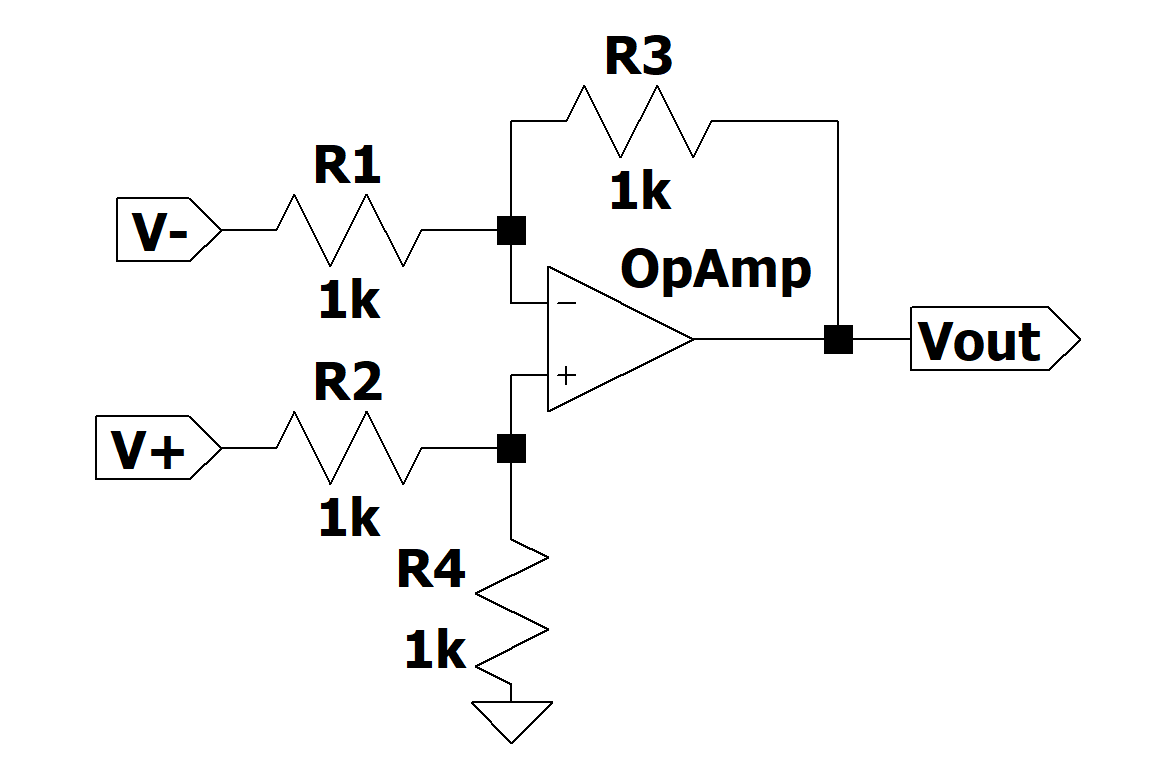}
	\caption{The subtractor circuit. Computes the function $V_{\text{out}}=V_+-V_-$.}
	\label{fig:subtractor}
\end{figure}

\begin{figure}[htbp]
	\centering
	\includegraphics[width = 0.6\textwidth]{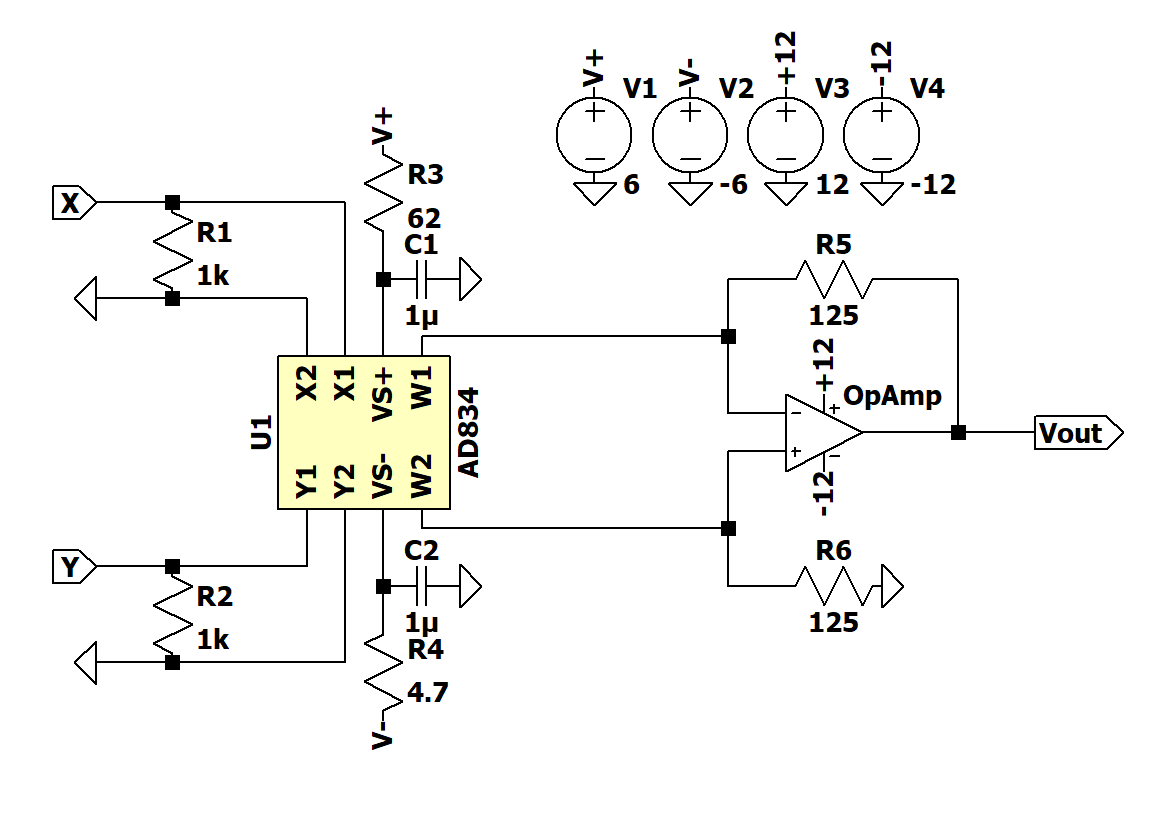}
	\caption{The multiplier circuit, constructed using the commercially available AD834 chip. It computes the function $V_{\text{out}}=XY / 1\text{V}$. }
	\label{fig:multiplier}
\end{figure}

Fig.~\ref{fig:adder}, \ref{fig:subtractor} illustrate the circuit designs for an adder and a subtractor, respectively. These are conventional designs that leverage the feedback loop of an op-amp. The circuit architecture for a multiplier is more intricate, and we have chosen to use the commercially available model, AD834, by Analog Devices \cite{AD834}. Figure~\ref{fig:multiplier} demonstrates the external circuit connections using the AD834 chip. This chip calculates the product of two input voltages, $X$ and $Y$, and the result is rendered as a current, $W$, according to the equation:
\begin{equation}
W = \frac{XY}{(1\text{V})^2}\times 4\text{mA}.
\end{equation}
We have employed an op-amp feedback loop to translate the output current back into a voltage signal. As a result, the complete transfer function of the multiplier module is given as:
\begin{equation}
V_{\text{out}}=\frac{XY}{1\text{V}}.
\end{equation}

\begin{figure}[htbp]
	\centering
	\includegraphics[width = 0.6\textwidth]{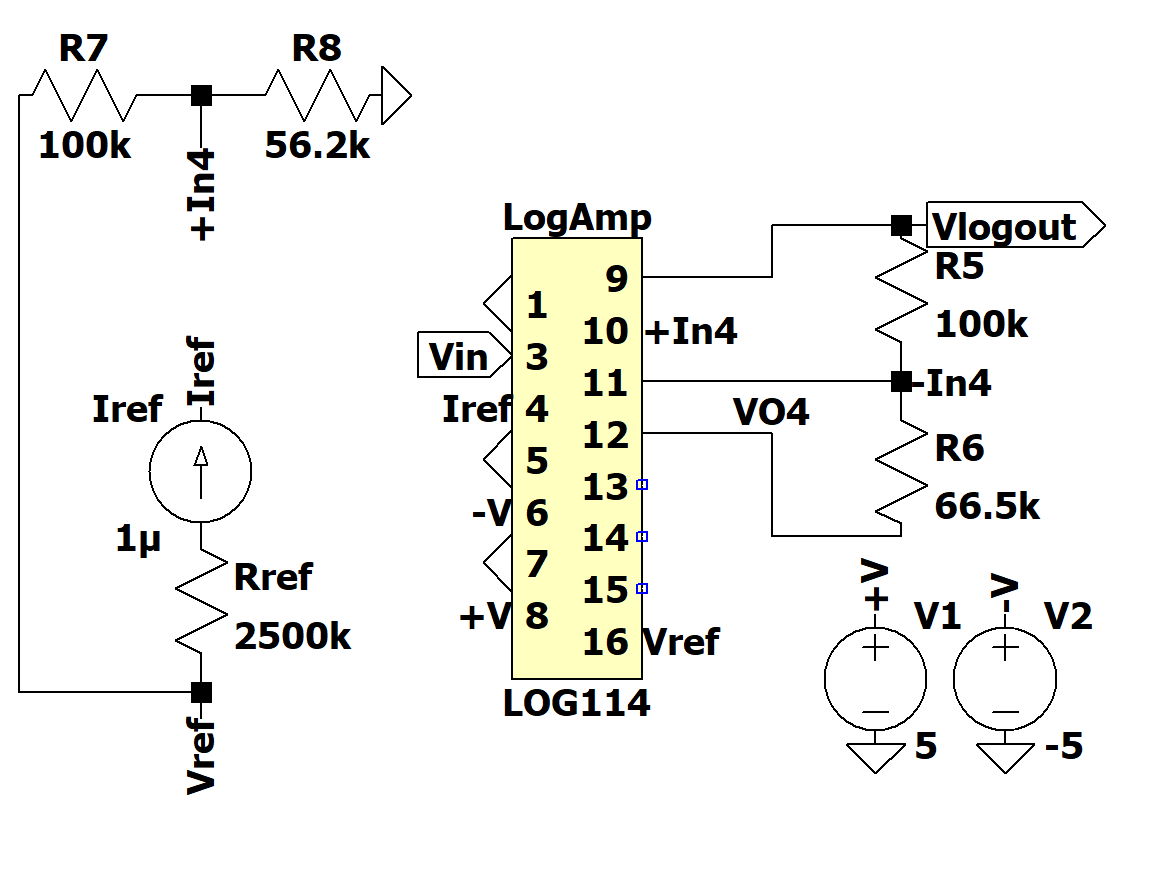}
	\caption{The logarithm amplifier, implemented using the commercially available design, the LOG114 chip. It computes the function $V_{\text{logout}}=(-0.375\text{V}) \log_{10}(V_{\text{in}}/1\mu \text{A})$.}
	\label{fig:log}
\end{figure}

Fig.~\ref{fig:log} showcases the design of the logarithm amplifier. This particular circuit computes the logarithm function as follows:
\begin{equation}
V_{\text{logout}}=(-0.375\text{V}) \log_{10}\frac{V_{\text{in}}}{1\mu \text{A}}.
\end{equation}
Again, we have made use of a commercially available model, LOG114, produced by Texas Instruments \cite{LOG114}. The circuit calculates the logarithm by utilizing the exponential relation between the emitter current and base voltage of a bipolar junction transistor (BJT). Two matching BJTs are used within this circuit design to effectively cancel out any temperature dependencies.

\begin{figure}[htbp]
	\centering
	\includegraphics[width = 0.6\textwidth]{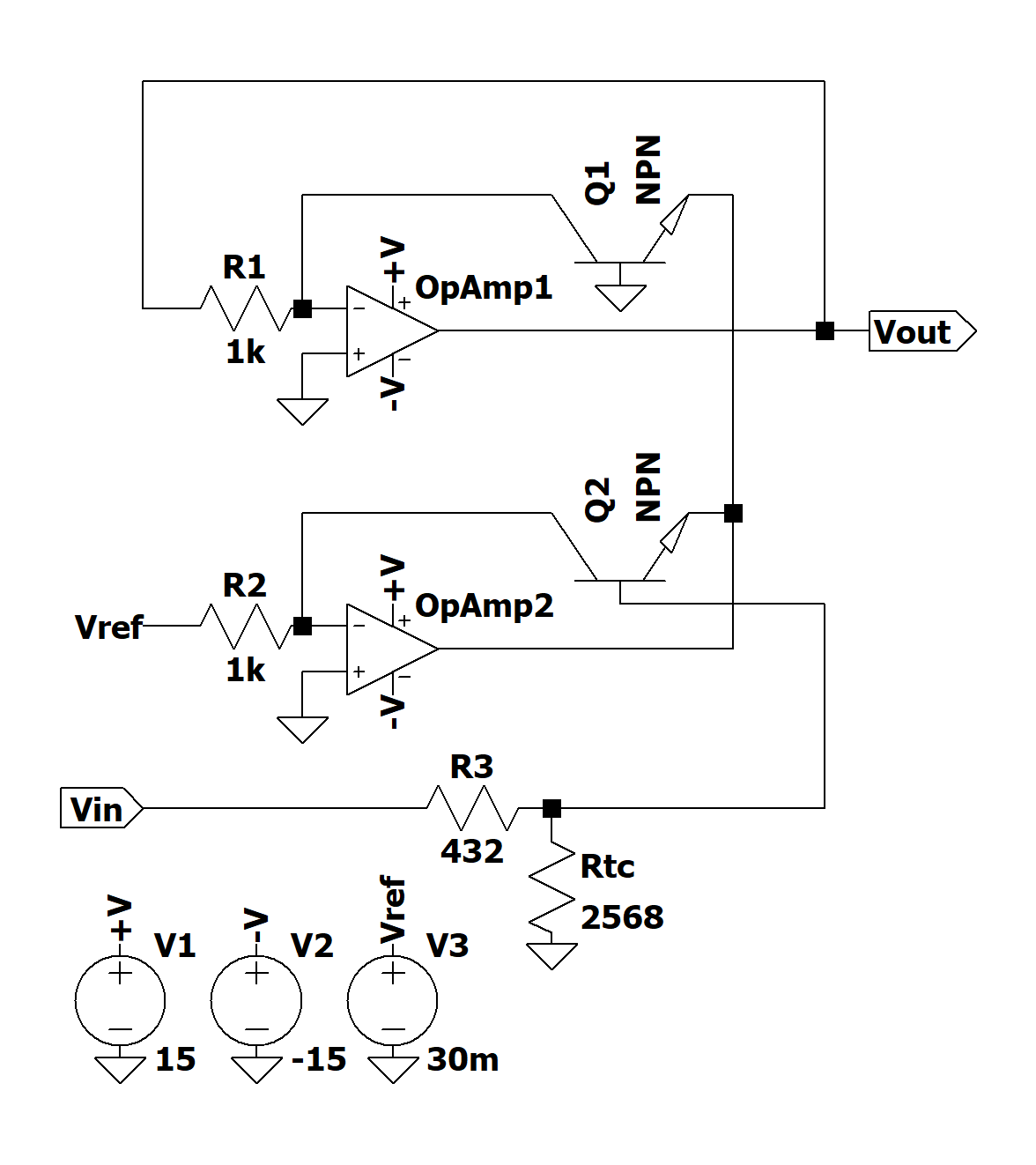}
	\caption{The anti-log amplifier. It computes the function $V_{\text{out}}=30\text{mV}\exp(-V_{\text{in}}/30\text{mV})$.}
	\label{fig:exp}
\end{figure}

Fig.~\ref{fig:exp} shows the design of the anti-log amplifier, which computes the exponential function, 
\begin{equation}
V_{\text{out}}=30\text{mV}\exp(-\frac{V_{\text{in}}}{30\text{mV}}).
\end{equation}
Due to the absence of commercially suitable designs for the anti-log amplifier, we had to implement this circuit ourselves. The final design bears similarity to the internal structure of the LOG114 circuit, but has been appropriately modified to accommodate for the exponential function.

\begin{figure}[htbp]
	\centering
	\includegraphics[width = 0.7\textwidth]{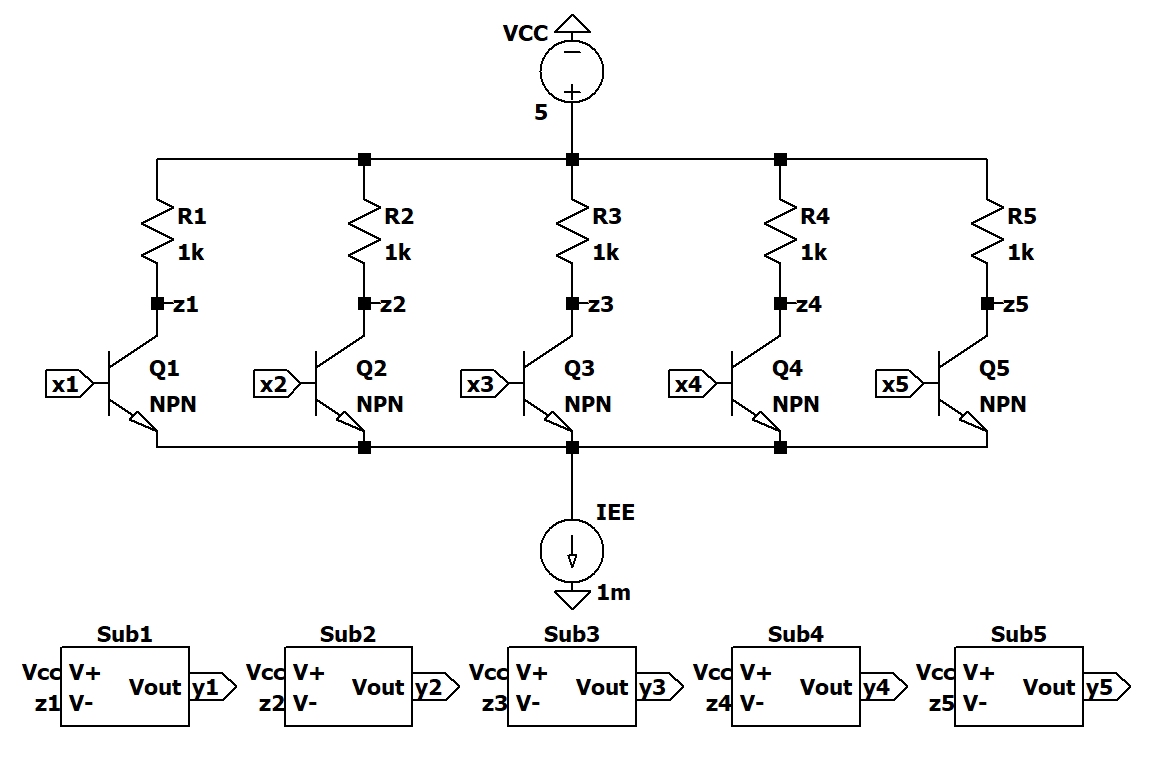}
	\caption{The circuit computing the softmax function. It outputs $y_i=(1V)\exp(x_i/V_T) / \sum_j \exp(x_j/V_T)$, where $V_T=k_B T/e$ is the thermal voltage. }
	\label{fig:softmax}
\end{figure}

Fig.~\ref{fig:softmax} presents a specially designed circuit for computing the softmax function, given by $\text{softmax}(x)_i=\frac{e^{x_i}}{\sum_j e^{x_j}}$. This circuit design builds upon pre-existing models \cite{elfadel1993softmax,sillman2023analog}. Rather than calculating multiple exponential functions, an array of BJTs with common emitters are employed. By setting a constant sum for the emitter currents, each base voltage regulates the distribution of current that flows through each respective BJT. The resulting expression precisely matches the softmax function. The current is then converted into a voltage measurement by observing the voltage drop across a resistor, after which a subtractor generates the desired final output. The transfer function of this circuit can be expressed as:
\begin{equation}
y_i=(1V)\exp(x_i/V_T) / \sum_j \exp(x_j/V_T)
\end{equation}
where $V_T=\frac{k_B T}{e}$ denotes the thermal voltage. Assuming a standard room temperature of $T=25^\circ \text{C}$, $V_T=25.68$mV. 

This circuit design permits the computation of the softmax function with an arbitrary number of inputs by incorporating additional BJTs configured in the same manner. It is crucial to note, however, that for the circuit to operate correctly, no current can flow into the $z_i$ ports of Fig.~\ref{fig:softmax}. This isolation can be achieved with the use of a voltage follower - an op-amp configured with unity gain - which does not affect the voltage output.

\begin{figure}[htbp]
	\centering
	\includegraphics[width = 0.6\textwidth]{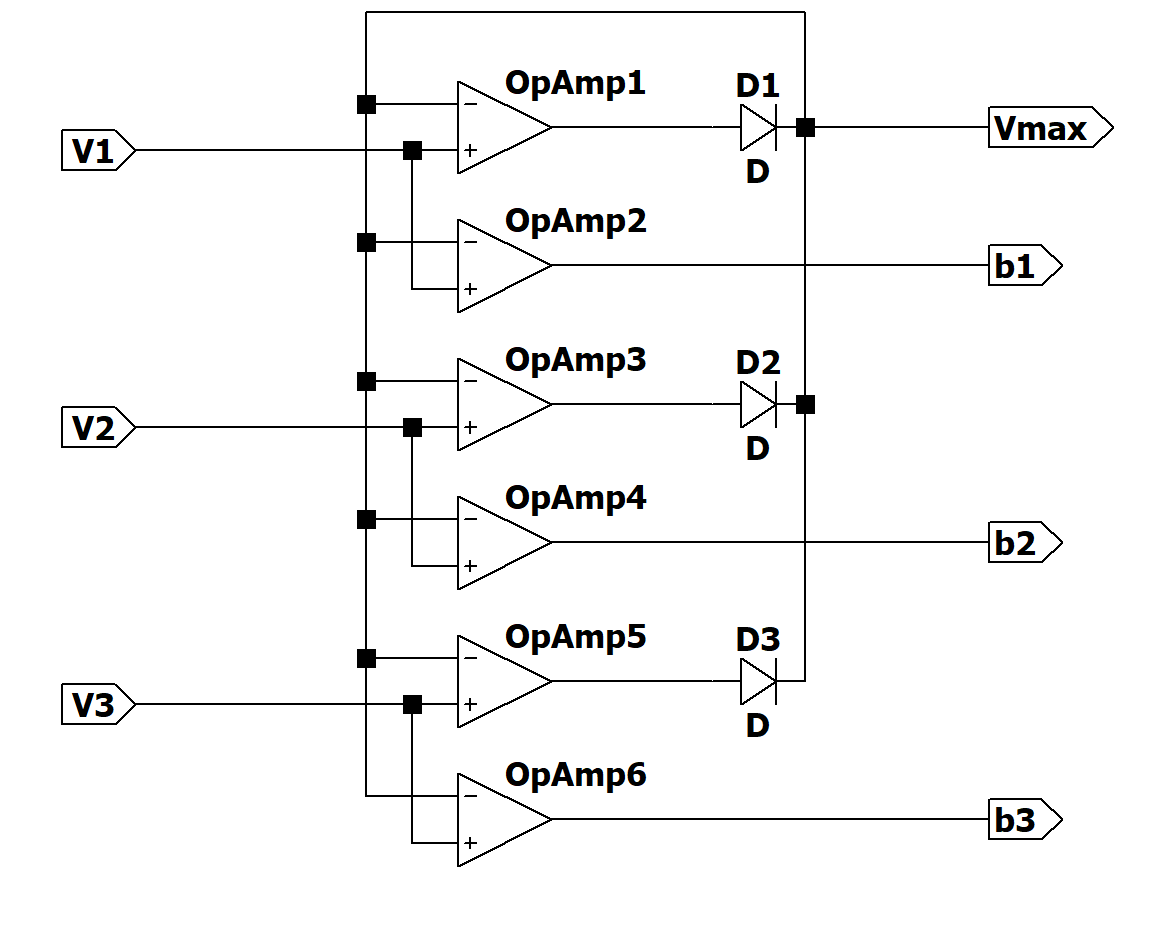}
	\caption{The comparator module. This circuit calculates $V_{\max} = \max(V_1, V_2, V_3)$, and the control signals $b_i$. If $V_i$ is the maximum, then $b_i=V_{\max}+V_D$, where $V_D$ is the voltage drop across the diodes. Otherwise, $b_i=-5$V.}
	\label{fig:comparator}
\end{figure}

To compute the maximum function, we employed the comparator module depicted in Fig.~\ref{fig:comparator}. This circuit is composed of two subcircuits. The feedback loop, which includes op-amps 1, 3, 5 and the three diodes, calculates the maximum voltage from the inputs $V_1, V_2$ and $V_3$, and outputs this value at the terminal $V_{\text{max}}$. This maximum voltage, $V_{\text{max}}$, is then fed back into three open-loop op-amps for comparison with the three input voltages. The results of these comparisons are output as $b_1, b_2$ and $b_3$, each of which indicates whether its corresponding input voltage is the maximum.

\begin{figure}[htbp]
	\centering
	\includegraphics[width = 0.55\textwidth]{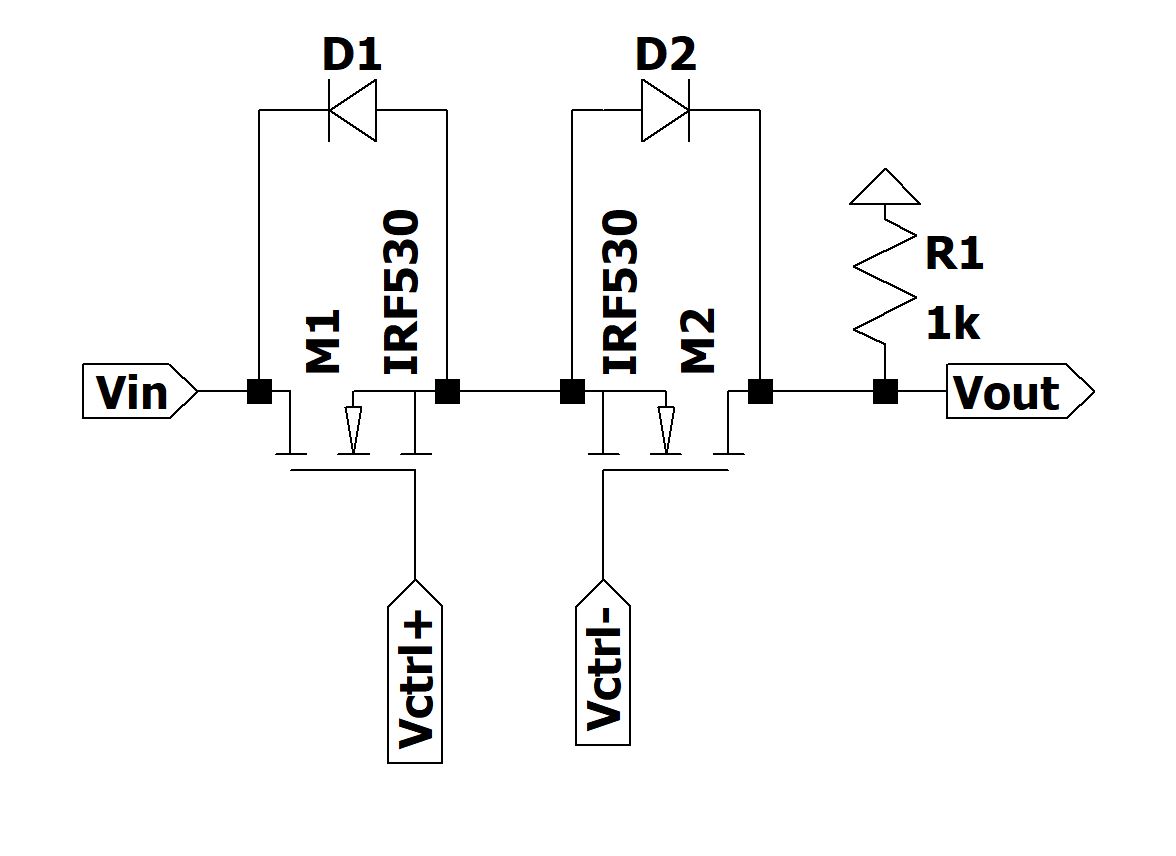}
	\caption{The bidirectional switch. It allows a positive signal to pass when $V_{\text{ctrl}+}$ is positive and a negative signal to pass when $V_{\text{ctrl}-}$ is positive. Conversely, it blocks a positive signal when $V_{\text{ctrl}+}$ is negative and a negative signal when $V_{\text{ctrl}-}$ is negative. The resistor $R_1$ facilitates the current flow when $V_{\text{out}}$ is connected to a large impedance.}
	\label{fig:switch}
\end{figure}

Fig.~\ref{fig:switch} illustrates the design of a bidirectional switch with two control signals. This mechanism modulates the current flow using the unidirectional conductivity properties of diodes and MOSFETs. Each MOSFET governs the signal flow in a specific direction. By independently managing the two opposing MOSFETs, we can selectively permit signals of certain polarities to pass through. This can be leveraged to set boundaries on a variable. For instance, positive currents to a capacitor can be blocked when its voltage surpasses the upper limit, while negative currents can be inhibited when the voltage falls below the lower threshold.

\section{Determining the optimal parameters}
\label{appendix:param}
The equations presented in Eqs.~\eqref{eq:v}-\eqref{eq:R} include seven constant parameters: $\alpha$, $\beta$, $\gamma$, $\delta$, $\epsilon$, $\zeta$, and $\eta$. These parameters can be fine-tuned to accelerate the numerical simulations. Following the guidelines from Ref.~\cite{bearden2020efficient}, we fixed $\alpha=5$, $\beta=20$, $\gamma=0.25$, $\delta=0.05$, and $\epsilon=10^{-3}$.

To identify the optimal values for $\zeta$, $\eta$, and the integration time step $\Delta t$, we employed Bayesian optimization, leveraging the Hyperopt library \cite{bergstra2013making}. Our process entailed solving 100 3-SAT instances, each comprising 1000 variables. We aimed at the maximization of the number of solved 3-SAT instances within a predefined step limit. The optimization process delivered the following optimal values: $\eta=3000$, $\zeta=3\times 10^{-3}$, and $\Delta t=0.14$.

However, in practice, the optimal values of these parameters may be problem-specific. To simplify the analysis, we vary one parameter at a time and scrutinize how the optimal parameter fluctuates in accordance with the number of variables in the problem.

\begin{figure}[htbp]
	\centering
	\includegraphics[width = 0.7\textwidth]{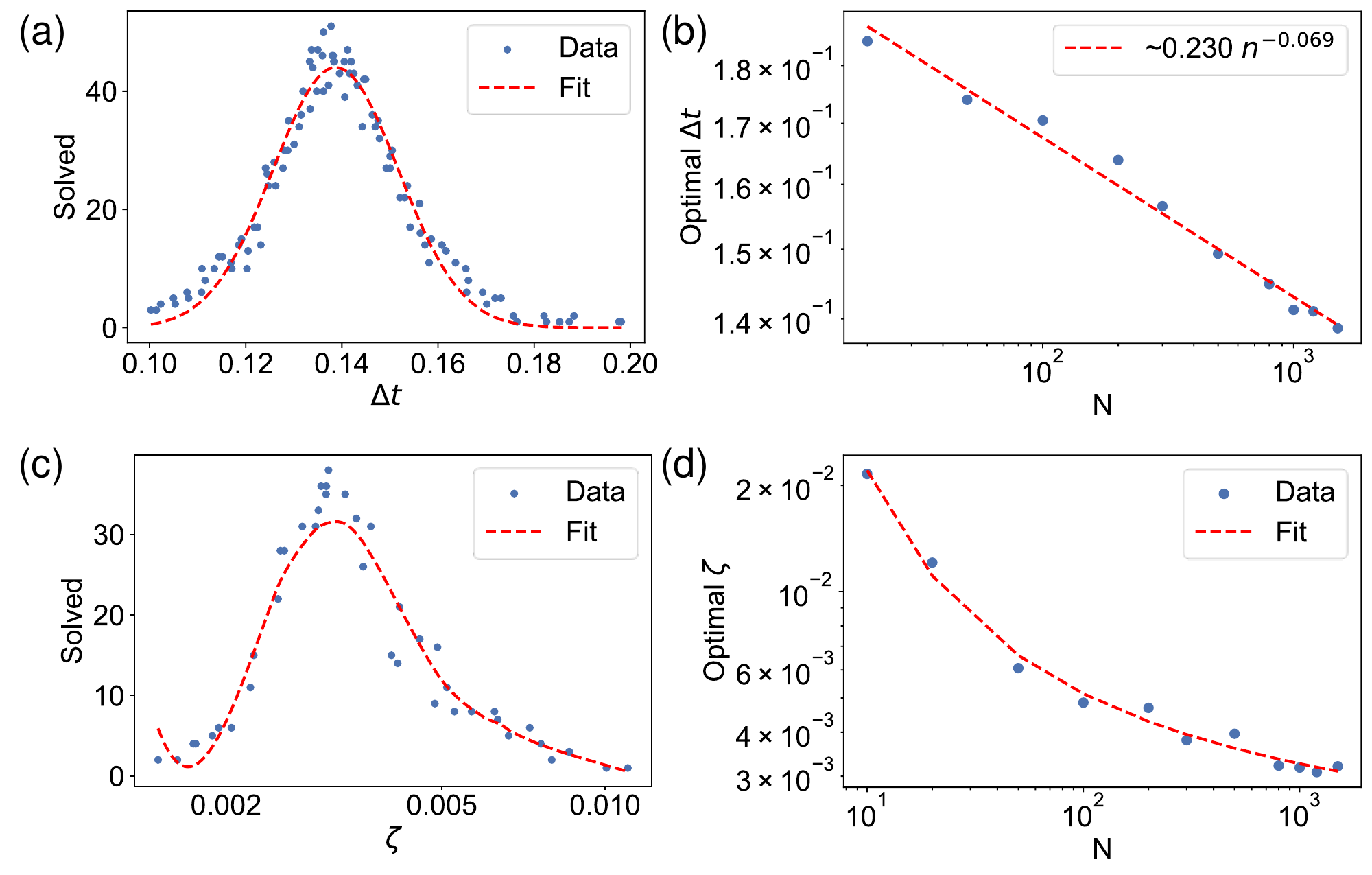}
	\caption{(a) The number of solved 3-SAT instances vs. $\Delta t$ within a given number of steps, for 100 3-SAT instances with 1500 variables. The optimal $\Delta t$ is chosen to be the peak of the fit. (b) The optimal $\Delta t$ as a function of the system size $N$: $\Delta t_{\text{optimal}}=0.230 N^{-0.069}$. (c) A similar analysis on the parameter $\zeta$. Again, 100 3-SAT instances with 1500 variables are repeatedly solved with different parameter $\zeta$. (d) The optimal $\zeta$ as a function of the system size $N$: $\ln \zeta_{\text{optimal}}=6.83(\ln N)^{-1.10}-6.53$.}
	\label{fig:param}
\end{figure}

Fig.~\ref{fig:param}(a) plots the number of successfully solved instances against $\Delta t$ for a set of 100 3-SAT problems, each composed of 1500 variables. The curve can be approximately fitted with a Gaussian, with the peak indicating the optimal $\Delta t$. By repeating this process for varying system sizes, we derive the relationship between the optimal $\Delta t$ and system size $N$. The results are demonstrated in Fig.~\ref{fig:param}(b). The optimal $\Delta t$ adheres to a power-law distribution, expressed as $\Delta t_{\text{optimal}}=0.230 N^{-0.069}$.

A similar analysis is performed for the parameter $\zeta$. Fig.~\ref{fig:param}(c) presents the results for $N=1500$. By replicating this process for various $N$, we obtain the optimal $\zeta$, as displayed in Fig.~\ref{fig:param}(d). This can be approximated with a polynomial in logarithmic space: $\ln \zeta_{\text{optimal}}=6.83(\ln N)^{-1.10}-6.53$. The parameter values derived from this analysis are applied to the numerical simulations discussed in the main text.

\section{Effects of physical imperfections}
\label{appendix:tolerance}

In this appendix, we explore the impact of physical imperfections, including component tolerances and capacitor leakage, on our system's ability to solve hard 3-SAT problems.

\begin{figure}[htbp]
	\centering
	\includegraphics[width = 0.7\textwidth]{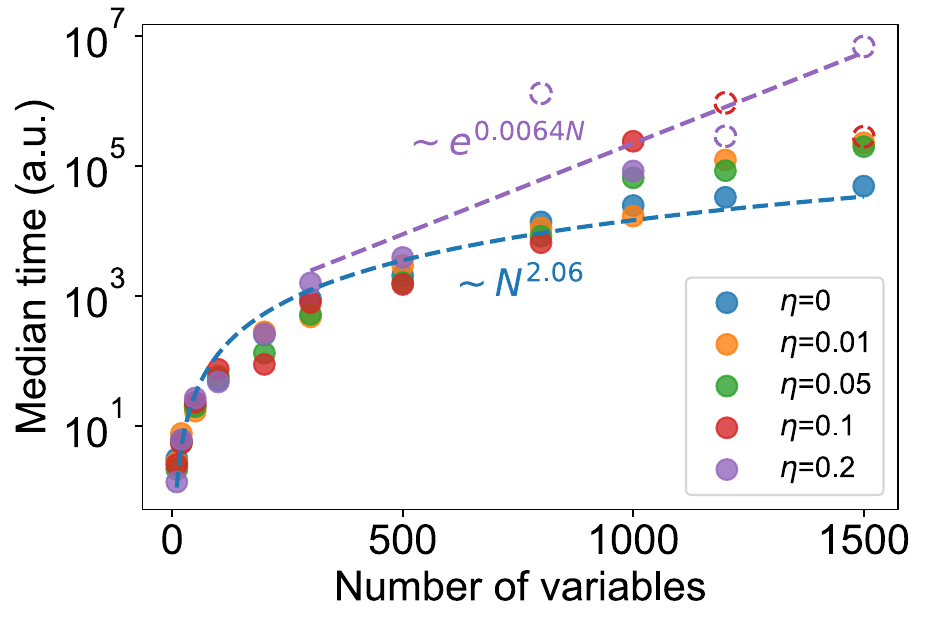}
	\caption{Extended analysis from Fig.~\ref{fig:tolerance}, illustrating the median integration time (in arbitrary units) as a function of the number of variables, $N$, for 3-SAT problems with a clause-to-variable ratio of 4.3. This simulation includes an error, $\eta$, in every addition and multiplication operation due to resistor tolerances, and accounts for a capacitor leakage current of $I_\mathrm{leak} = 10^{-1}\mu\mathrm{A}/(\mu\mathrm{F} \cdot \mathrm{V}) \cdot CV$. Solid data points represent conditions where at least 51 out of 100 distinct 3-SAT instances are solved, allowing direct calculation of the median integration time. Dashed circles indicate estimated median times for scenarios where fewer than half of the instances are resolved. The findings underscore that while capacitor leakage minimally affects the dynamics, resistor tolerances notably impact system behavior. Noise levels, $\eta$, slightly influence solution times for smaller instances ($N \leq 500$); however, as expected, for larger $N$, the median solution time shows exponential growth, at much higher $\eta$ values.}
	\label{fig:tolerance_appendix}
\end{figure}

In the results section of the main text, we discussed how various physical factors affect performance and demonstrated our system's tolerance to low noise levels in resistor values in Fig.~\ref{fig:tolerance}. Here, we extend this analysis with additional scaling curves in Fig.\ref{fig:tolerance_appendix}, examining noise levels, $\eta$, up to 0.2 and variable counts, $N$, up to 1500. The experimental setup remains consistent with that described in Fig.~\ref{fig:tolerance}, unless specified otherwise.

As $\eta$ and $N$ increase, we observe a significant growth in solution time. We evaluated 100 distinct 3-SAT problems for each size. To manage computational costs, simulations were terminated after $2 \times 10^6$ integration steps. If more than half of the instances are solved within this limit, the median solution time is directly calculated. If fewer than half are solved, we estimate the median solution time based on the number of solved instances at $2\times 10^6$ integration steps, assuming a uniform distribution of solution times across all instances. 

Fig.~\ref{fig:tolerance_appendix} presents the median integration time as a function of $N$ and $\eta$. Estimated medians for cases where fewer than half of the instances are solved are depicted with dashed circles. Remarkably, even at $\eta$ values as high as 0.2—which could potentially double or halve outcomes in the computed derivatives—our system reliably solves hard 3-SAT problems for $N \leq 500$. Beyond this threshold, however, the median solution time begins to show exponential growth for large $\eta$. Despite this, for lower $\eta$ values, our results consistently demonstrate effective resolution of these challenging problems, with the median solution time displaying a quadratic scaling for $N$ up to 1500.

\end{appendices}
\FloatBarrier
\bibliography{bibliography}


\begin{thebibliography}{41}
\ifx \bisbn   \undefined \def \bisbn  #1{ISBN #1}\fi
\ifx \binits  \undefined \def \binits#1{#1}\fi
\ifx \bauthor  \undefined \def \bauthor#1{#1}\fi
\ifx \batitle  \undefined \def \batitle#1{#1}\fi
\ifx \bjtitle  \undefined \def \bjtitle#1{#1}\fi
\ifx \bvolume  \undefined \def \bvolume#1{\textbf{#1}}\fi
\ifx \byear  \undefined \def \byear#1{#1}\fi
\ifx \bissue  \undefined \def \bissue#1{#1}\fi
\ifx \bfpage  \undefined \def \bfpage#1{#1}\fi
\ifx \blpage  \undefined \def \blpage #1{#1}\fi
\ifx \burl  \undefined \def \burl#1{\textsf{#1}}\fi
\ifx \doiurl  \undefined \def \doiurl#1{\url{https://doi.org/#1}}\fi
\ifx \betal  \undefined \def \betal{\textit{et al.}}\fi
\ifx \binstitute  \undefined \def \binstitute#1{#1}\fi
\ifx \binstitutionaled  \undefined \def \binstitutionaled#1{#1}\fi
\ifx \bctitle  \undefined \def \bctitle#1{#1}\fi
\ifx \beditor  \undefined \def \beditor#1{#1}\fi
\ifx \bpublisher  \undefined \def \bpublisher#1{#1}\fi
\ifx \bbtitle  \undefined \def \bbtitle#1{#1}\fi
\ifx \bedition  \undefined \def \bedition#1{#1}\fi
\ifx \bseriesno  \undefined \def \bseriesno#1{#1}\fi
\ifx \blocation  \undefined \def \blocation#1{#1}\fi
\ifx \bsertitle  \undefined \def \bsertitle#1{#1}\fi
\ifx \bsnm \undefined \def \bsnm#1{#1}\fi
\ifx \bsuffix \undefined \def \bsuffix#1{#1}\fi
\ifx \bparticle \undefined \def \bparticle#1{#1}\fi
\ifx \barticle \undefined \def \barticle#1{#1}\fi
\bibcommenthead
\ifx \bconfdate \undefined \def \bconfdate #1{#1}\fi
\ifx \botherref \undefined \def \botherref #1{#1}\fi
\ifx \url \undefined \def \url#1{\textsf{#1}}\fi
\ifx \bchapter \undefined \def \bchapter#1{#1}\fi
\ifx \bbook \undefined \def \bbook#1{#1}\fi
\ifx \bcomment \undefined \def \bcomment#1{#1}\fi
\ifx \oauthor \undefined \def \oauthor#1{#1}\fi
\ifx \citeauthoryear \undefined \def \citeauthoryear#1{#1}\fi
\ifx \endbibitem  \undefined \def \endbibitem {}\fi
\ifx \bconflocation  \undefined \def \bconflocation#1{#1}\fi
\ifx \arxivurl  \undefined \def \arxivurl#1{\textsf{#1}}\fi
\csname PreBibitemsHook\endcsname

\bibitem[\protect\citeauthoryear{Schaller}{1997}]{schaller1997moore}
\begin{barticle}
\bauthor{\bsnm{Schaller}, \binits{R.R.}}:
\batitle{Moore's law: past, present and future}.
\bjtitle{IEEE spectrum}
\bvolume{34}(\bissue{6}),
\bfpage{52}--\blpage{59}
(\byear{1997})
\end{barticle}
\endbibitem

\bibitem[\protect\citeauthoryear{Shalf}{2020}]{shalf2020future}
\begin{barticle}
\bauthor{\bsnm{Shalf}, \binits{J.}}:
\batitle{The future of computing beyond moore’s law}.
\bjtitle{Philosophical Transactions of the Royal Society A}
\bvolume{378}(\bissue{2166}),
\bfpage{20190061}
(\byear{2020})
\end{barticle}
\endbibitem

\bibitem[\protect\citeauthoryear{Nielsen and Chuang}{2010}]{nielsen2010quantum}
\begin{bbook}
\bauthor{\bsnm{Nielsen}, \binits{M.A.}},
\bauthor{\bsnm{Chuang}, \binits{I.L.}}:
\bbtitle{Quantum Computation and Quantum Information}.
\bpublisher{Cambridge university press},
\blocation{Cambridge, United Kingdom}
(\byear{2010})
\end{bbook}
\endbibitem

\bibitem[\protect\citeauthoryear{Islam et~al.}{2019}]{Wong}
\begin{barticle}
\bauthor{\bsnm{Islam}, \binits{R.}},
\bauthor{\bsnm{Li}, \binits{H.}},
\bauthor{\bsnm{Chen}, \binits{P.-Y.}},
\bauthor{\bsnm{Wan}, \binits{W.}},
\bauthor{\bsnm{Chen}, \binits{H.-Y.}},
\bauthor{\bsnm{Gao}, \binits{B.}},
\bauthor{\bsnm{Wu}, \binits{H.}},
\bauthor{\bsnm{Yu}, \binits{S.}},
\bauthor{\bsnm{Saraswat}, \binits{K.}},
\bauthor{\bsnm{Wong}, \binits{H.-S.P.}}:
\batitle{Device and materials requirements for neuromorphic computing}.
\bjtitle{Journal of Physics D: Applied Physics}
\bvolume{52},
\bfpage{113001}
(\byear{2019})
\end{barticle}
\endbibitem

\bibitem[\protect\citeauthoryear{Solli and Jalali}{2015}]{solli2015analog}
\begin{barticle}
\bauthor{\bsnm{Solli}, \binits{D.R.}},
\bauthor{\bsnm{Jalali}, \binits{B.}}:
\batitle{Analog optical computing}.
\bjtitle{Nature Photonics}
\bvolume{9}(\bissue{11}),
\bfpage{704}--\blpage{706}
(\byear{2015})
\end{barticle}
\endbibitem

\bibitem[\protect\citeauthoryear{De~Silva and Uchiyama}{2007}]{de2007molecular}
\begin{barticle}
\bauthor{\bsnm{De~Silva}, \binits{A.P.}},
\bauthor{\bsnm{Uchiyama}, \binits{S.}}:
\batitle{Molecular logic and computing}.
\bjtitle{Nature nanotechnology}
\bvolume{2}(\bissue{7}),
\bfpage{399}--\blpage{410}
(\byear{2007})
\end{barticle}
\endbibitem

\bibitem[\protect\citeauthoryear{Shor}{1999}]{shor1999polynomial}
\begin{barticle}
\bauthor{\bsnm{Shor}, \binits{P.W.}}:
\batitle{Polynomial-time algorithms for prime factorization and discrete logarithms on a quantum computer}.
\bjtitle{SIAM review}
\bvolume{41}(\bissue{2}),
\bfpage{303}--\blpage{332}
(\byear{1999})
\end{barticle}
\endbibitem

\bibitem[\protect\citeauthoryear{{Di Ventra} and Pershin}{2013}]{diventra13a}
\begin{barticle}
\bauthor{\bsnm{{Di Ventra}}, \binits{M.}},
\bauthor{\bsnm{Pershin}, \binits{Y.V.}}:
\batitle{The parallel approach}.
\bjtitle{Nature Physics}
\bvolume{9},
\bfpage{200}
(\byear{2013})
\end{barticle}
\endbibitem

\bibitem[\protect\citeauthoryear{Traversa and {Di Ventra}}{2015}]{traversa2015universal}
\begin{barticle}
\bauthor{\bsnm{Traversa}, \binits{F.L.}},
\bauthor{\bsnm{{Di Ventra}}, \binits{M.}}:
\batitle{Universal memcomputing machines}.
\bjtitle{IEEE transactions on neural networks and learning systems}
\bvolume{26}(\bissue{11}),
\bfpage{2702}--\blpage{2715}
(\byear{2015})
\end{barticle}
\endbibitem

\bibitem[\protect\citeauthoryear{{Di Ventra}}{2022}]{di2022memcomputing}
\begin{bbook}
\bauthor{\bsnm{{Di Ventra}}, \binits{M.}}:
\bbtitle{MemComputing: Fundamentals and Applications}.
\bpublisher{Oxford University Press},
\blocation{Oxford}
(\byear{2022})
\end{bbook}
\endbibitem

\bibitem[\protect\citeauthoryear{Traversa and Di~Ventra}{2017}]{traversa2017polynomial}
\begin{botherref}
\oauthor{\bsnm{Traversa}, \binits{F.L.}},
\oauthor{\bsnm{Di~Ventra}, \binits{M.}}:
Polynomial-time solution of prime factorization and np-complete problems with digital memcomputing machines.
Chaos: An Interdisciplinary Journal of Nonlinear Science
\textbf{27}(2)
(2017)
\end{botherref}
\endbibitem

\bibitem[\protect\citeauthoryear{{Di Ventra} and Pershin}{2023}]{di2023memory}
\begin{bbook}
\bauthor{\bsnm{{Di Ventra}}, \binits{M.}},
\bauthor{\bsnm{Pershin}, \binits{Y.V.}}:
\bbtitle{Memristors and Memelements: Mathematics, Physics, and Fiction}.
\bpublisher{Springer},
\blocation{Berlin, Germany}
(\byear{2023})
\end{bbook}
\endbibitem

\bibitem[\protect\citeauthoryear{Di~Ventra et~al.}{2017}]{di2017topological}
\begin{barticle}
\bauthor{\bsnm{Di~Ventra}, \binits{M.}},
\bauthor{\bsnm{Traversa}, \binits{F.L.}},
\bauthor{\bsnm{Ovchinnikov}, \binits{I.V.}}:
\batitle{Topological field theory and computing with instantons}.
\bjtitle{Annalen der Physik}
\bvolume{529}(\bissue{12}),
\bfpage{1700123}
(\byear{2017})
\end{barticle}
\endbibitem

\bibitem[\protect\citeauthoryear{Di~Ventra and Ovchinnikov}{2019}]{di2019digital}
\begin{barticle}
\bauthor{\bsnm{Di~Ventra}, \binits{M.}},
\bauthor{\bsnm{Ovchinnikov}, \binits{I.V.}}:
\batitle{Digital memcomputing: from logic to dynamics to topology}.
\bjtitle{Annals of Physics}
\bvolume{409},
\bfpage{167935}
(\byear{2019})
\end{barticle}
\endbibitem

\bibitem[\protect\citeauthoryear{Bearden et~al.}{2020}]{bearden2020efficient}
\begin{barticle}
\bauthor{\bsnm{Bearden}, \binits{S.R.}},
\bauthor{\bsnm{Pei}, \binits{Y.R.}},
\bauthor{\bsnm{Di~Ventra}, \binits{M.}}:
\batitle{Efficient solution of boolean satisfiability problems with digital memcomputing}.
\bjtitle{Scientific reports}
\bvolume{10}(\bissue{1}),
\bfpage{19741}
(\byear{2020})
\end{barticle}
\endbibitem

\bibitem[\protect\citeauthoryear{Di~Ventra and Traversa}{2017a}]{di2017chaos}
\begin{barticle}
\bauthor{\bsnm{Di~Ventra}, \binits{M.}},
\bauthor{\bsnm{Traversa}, \binits{F.L.}}:
\batitle{Absence of chaos in digital memcomputing machines with solutions}.
\bjtitle{Physics Letters A}
\bvolume{381}(\bissue{38}),
\bfpage{3255}--\blpage{3257}
(\byear{2017})
\end{barticle}
\endbibitem

\bibitem[\protect\citeauthoryear{Di~Ventra and Traversa}{2017b}]{di2017periodic}
\begin{botherref}
\oauthor{\bsnm{Di~Ventra}, \binits{M.}},
\oauthor{\bsnm{Traversa}, \binits{F.L.}}:
Absence of periodic orbits in digital memcomputing machines with solutions.
Chaos: An Interdisciplinary Journal of Nonlinear Science
\textbf{27}(10)
(2017)
\end{botherref}
\endbibitem

\bibitem[\protect\citeauthoryear{Traversa et~al.}{2018}]{traversa2018evidence}
\begin{botherref}
\oauthor{\bsnm{Traversa}, \binits{F.L.}},
\oauthor{\bsnm{Cicotti}, \binits{P.}},
\oauthor{\bsnm{Sheldon}, \binits{F.}},
\oauthor{\bsnm{Di~Ventra}, \binits{M.}}:
Evidence of exponential speed-up in the solution of hard optimization problems.
Complexity
\textbf{2018}
(2018)
\end{botherref}
\endbibitem

\bibitem[\protect\citeauthoryear{Traversa and Di~Ventra}{2018}]{traversa2018memcomputing}
\begin{botherref}
\oauthor{\bsnm{Traversa}, \binits{F.L.}},
\oauthor{\bsnm{Di~Ventra}, \binits{M.}}:
Memcomputing integer linear programming.
arXiv preprint arXiv:1808.09999
(2018)
\end{botherref}
\endbibitem

\bibitem[\protect\citeauthoryear{Sheldon et~al.}{2019}]{sheldon2019stress}
\begin{barticle}
\bauthor{\bsnm{Sheldon}, \binits{F.}},
\bauthor{\bsnm{Cicotti}, \binits{P.}},
\bauthor{\bsnm{Traversa}, \binits{F.L.}},
\bauthor{\bsnm{Di~Ventra}, \binits{M.}}:
\batitle{Stress-testing memcomputing on hard combinatorial optimization problems}.
\bjtitle{IEEE transactions on neural networks and learning systems}
\bvolume{31}(\bissue{6}),
\bfpage{2222}--\blpage{2226}
(\byear{2019})
\end{barticle}
\endbibitem

\bibitem[\protect\citeauthoryear{}{}]{Company}
\begin{botherref}
\url{https://www.memcpu.com/}
\end{botherref}
\endbibitem

\bibitem[\protect\citeauthoryear{Sharp et~al.}{2023}]{sharp2023scaling}
\begin{botherref}
\oauthor{\bsnm{Sharp}, \binits{T.}},
\oauthor{\bsnm{Khare}, \binits{R.}},
\oauthor{\bsnm{Pederson}, \binits{E.}},
\oauthor{\bsnm{Traversa}, \binits{F.L.}}:
Scaling up prime factorization with self-organizing gates: A memcomputing approach.
arXiv preprint arXiv:2309.08198
(2023)
{\href{https://arxiv.org/abs/2309.08198}{{arXiv:2309.08198}}}
{[cs.ET]}
\end{botherref}
\endbibitem

\bibitem[\protect\citeauthoryear{Rivest et~al.}{1978}]{RSA}
\begin{barticle}
\bauthor{\bsnm{Rivest}, \binits{R.L.}},
\bauthor{\bsnm{Shamir}, \binits{A.}},
\bauthor{\bsnm{Adleman}, \binits{L.}}:
\batitle{A method for obtaining digital signatures and public-key cryptosystems}.
\bjtitle{Communications of the ACM}
\bvolume{21}(\bissue{2}),
\bfpage{120}--\blpage{126}
(\byear{1978})
\end{barticle}
\endbibitem

\bibitem[\protect\citeauthoryear{Stuart}{1994}]{stuart1994numerical}
\begin{barticle}
\bauthor{\bsnm{Stuart}, \binits{A.M.}}:
\batitle{Numerical analysis of dynamical systems}.
\bjtitle{Acta numerica}
\bvolume{3},
\bfpage{467}--\blpage{572}
(\byear{1994})
\end{barticle}
\endbibitem

\bibitem[\protect\citeauthoryear{Zhang and Di~Ventra}{2021}]{zhang2021directed}
\begin{barticle}
\bauthor{\bsnm{Zhang}, \binits{Y.-H.}},
\bauthor{\bsnm{Di~Ventra}, \binits{M.}}:
\batitle{Directed percolation and numerical stability of simulations of digital memcomputing machines}.
\bjtitle{Chaos: An Interdisciplinary Journal of Nonlinear Science}
\bvolume{31},
\bfpage{063127}
(\byear{2021})
\end{barticle}
\endbibitem

\bibitem[\protect\citeauthoryear{Gypens et~al.}{2021}]{Pinna}
\begin{barticle}
\bauthor{\bsnm{Gypens}, \binits{P.}},
\bauthor{\bsnm{Leliaert}, \binits{J.}},
\bauthor{\bsnm{{Di Ventra}}, \binits{M.}},
\bauthor{\bsnm{Van~Waeyenberge}, \binits{B.}},
\bauthor{\bsnm{Pinna}, \binits{D.}}:
\batitle{Nanomagnetic self-organizing logic gates}.
\bjtitle{Physical Review Applied}
\bvolume{16},
\bfpage{024055}
(\byear{2021})
\end{barticle}
\endbibitem

\bibitem[\protect\citeauthoryear{Nguyen et~al.}{2023}]{nguyen2023hardware}
\begin{botherref}
\oauthor{\bsnm{Nguyen}, \binits{D.C.}},
\oauthor{\bsnm{Zhang}, \binits{Y.-H.}},
\oauthor{\bsnm{Di~Ventra}, \binits{M.}},
\oauthor{\bsnm{Pershin}, \binits{Y.V.}}:
Hardware implementation of digital memcomputing on small-size fpgas.
arXiv preprint arXiv:2305.01061
(2023)
\end{botherref}
\endbibitem

\bibitem[\protect\citeauthoryear{Cook}{1971}]{cook1971complexity}
\begin{bchapter}
\bauthor{\bsnm{Cook}, \binits{S.A.}}:
\bctitle{The complexity of theorem-proving procedures}.
In: \bbtitle{Proceedings of the Third Annual ACM Symposium on Theory of Computing}.
\bsertitle{STOC '71},
pp. \bfpage{151}--\blpage{158}.
\bpublisher{Association for Computing Machinery},
\blocation{New York, NY, USA}
(\byear{1971}).
\doiurl{10.1145/800157.805047} .
\burl{https://doi.org/10.1145/800157.805047}
\end{bchapter}
\endbibitem

\bibitem[\protect\citeauthoryear{Primosch et~al.}{2023}]{primosch2023self}
\begin{barticle}
\bauthor{\bsnm{Primosch}, \binits{D.}},
\bauthor{\bsnm{Zhang}, \binits{Y.-H.}},
\bauthor{\bsnm{{Di Ventra}}, \binits{M.}}:
\batitle{Self-averaging of digital memcomputing machines}.
\bjtitle{Phys. Rev. E}
\bvolume{108},
\bfpage{034306}
(\byear{2023})
\end{barticle}
\endbibitem

\bibitem[\protect\citeauthoryear{Eggersglüß and Drechsler}{2010}]{eggersgluss2010robust}
\begin{bchapter}
\bauthor{\bsnm{Eggersglüß}, \binits{S.}},
\bauthor{\bsnm{Drechsler}, \binits{R.}}:
\bctitle{Robust algorithms for high quality test pattern generation using boolean satisfiability}.
In: \bbtitle{2010 IEEE International Test Conference},
\bconflocation{Austin, TX, USA},
pp. \bfpage{1}--\blpage{10}
(\byear{2010}).
\doiurl{10.1109/TEST.2010.5699289}
\end{bchapter}
\endbibitem

\bibitem[\protect\citeauthoryear{Gomes et~al.}{2000}]{gomes2000heavy}
\begin{barticle}
\bauthor{\bsnm{Gomes}, \binits{C.P.}},
\bauthor{\bsnm{Selman}, \binits{B.}},
\bauthor{\bsnm{Crato}, \binits{N.}},
\bauthor{\bsnm{Kautz}, \binits{H.}}:
\batitle{Heavy-tailed phenomena in satisfiability and constraint satisfaction problems}.
\bjtitle{Journal of automated reasoning}
\bvolume{24}(\bissue{1-2}),
\bfpage{67}--\blpage{100}
(\byear{2000})
\end{barticle}
\endbibitem

\bibitem[\protect\citeauthoryear{Vladimirescu}{1994}]{vladimirescu1994spice}
\begin{bbook}
\bauthor{\bsnm{Vladimirescu}, \binits{A.}}:
\bbtitle{The SPICE Book}.
\bpublisher{Wiley},
\blocation{New York}
(\byear{1994})
\end{bbook}
\endbibitem

\bibitem[\protect\citeauthoryear{Kundert}{2006}]{kundert2006designer}
\begin{bbook}
\bauthor{\bsnm{Kundert}, \binits{K.}}:
\bbtitle{The Designer’s Guide to SPICE and SPECTRE{\textregistered}}.
\bpublisher{Springer},
\blocation{Berlin, Germany}
(\byear{2006})
\end{bbook}
\endbibitem

\bibitem[\protect\citeauthoryear{}{}]{github_link}
\begin{botherref}
\url{https://github.com/yuanhangzhang98/DMM_circuit}
\end{botherref}
\endbibitem

\bibitem[\protect\citeauthoryear{Barthel et~al.}{2002}]{barthel2002hiding}
\begin{barticle}
\bauthor{\bsnm{Barthel}, \binits{W.}},
\bauthor{\bsnm{Hartmann}, \binits{A.K.}},
\bauthor{\bsnm{Leone}, \binits{M.}},
\bauthor{\bsnm{Ricci-Tersenghi}, \binits{F.}},
\bauthor{\bsnm{Weigt}, \binits{M.}},
\bauthor{\bsnm{Zecchina}, \binits{R.}}:
\batitle{Hiding solutions in random satisfiability problems: A statistical mechanics approach}.
\bjtitle{Physical review letters}
\bvolume{88}(\bissue{18}),
\bfpage{188701}
(\byear{2002})
\end{barticle}
\endbibitem

\bibitem[\protect\citeauthoryear{International Electrotechnical Commission}{2016}]{IEC60384-4}
\begin{botherref}
International Electrotechnical Commission:
IEC 60384-4, Fixed Capacitors for Use in Electronic Equipment - Part 4: Sectional Specification - Fixed Aluminium Electrolytic Capacitors with Solid (MnO2) and Non-solid Electrolyte.
(2016).
International Electrotechnical Commission.
\url{https://webstore.iec.ch/publication/25648}
\end{botherref}
\endbibitem

\bibitem[\protect\citeauthoryear{Analog Devices}{2012}]{AD834}
\begin{botherref}
Analog Devices:
AD834, 500 MHz Four-Quadrant Multiplier.
(2012).
Analog Devices. Rev. F.
\url{https://www.analog.com/media/en/technical-documentation/data-sheets/AD834.pdf}
\end{botherref}
\endbibitem

\bibitem[\protect\citeauthoryear{Texas Instruments}{2004}]{LOG114}
\begin{botherref}
Texas Instruments:
Single-Supply, High-Speed, Precision Logarithmic Amplifier Datasheet.
(2004).
Texas Instruments. Rev. A.
\url{https://www.ti.com/lit/ds/symlink/log114.pdf}
\end{botherref}
\endbibitem

\bibitem[\protect\citeauthoryear{Elfadel and Wyatt~Jr}{1993}]{elfadel1993softmax}
\begin{botherref}
\oauthor{\bsnm{Elfadel}, \binits{I.M.}},
\oauthor{\bsnm{Wyatt~Jr}, \binits{J.L.}}:
The" softmax" nonlinearity: Derivation using statistical mechanics and useful properties as a multiterminal analog circuit element.
Advances in neural information processing systems
\textbf{6}
(1993)
\end{botherref}
\endbibitem

\bibitem[\protect\citeauthoryear{Sillman}{2023}]{sillman2023analog}
\begin{botherref}
\oauthor{\bsnm{Sillman}, \binits{J.}}:
Analog implementation of the softmax function.
arXiv preprint arXiv:2305.13649
(2023)
\end{botherref}
\endbibitem

\bibitem[\protect\citeauthoryear{Bergstra et~al.}{2013}]{bergstra2013making}
\begin{bchapter}
\bauthor{\bsnm{Bergstra}, \binits{J.}},
\bauthor{\bsnm{Yamins}, \binits{D.}},
\bauthor{\bsnm{Cox}, \binits{D.}}:
\bctitle{Making a science of model search: Hyperparameter optimization in hundreds of dimensions for vision architectures}.
In: \bbtitle{International Conference on Machine Learning},
\bconflocation{Atlanta, Georgia, USA},
pp. \bfpage{115}--\blpage{123}
(\byear{2013}).
\bcomment{PMLR}
\end{bchapter}
\endbibitem

\end{thebibliography}
\end{document}